 \documentclass[final,5p,times,twocolumn,authoryear]{elsarticle}

\usepackage{amssymb}
\usepackage{amsmath}
\usepackage{lipsum}
\usepackage{float}
\usepackage{xcolor}
\usepackage{tabularx}

\usepackage[utf8]{inputenc}
\usepackage{graphicx}
\usepackage{cuted}
\usepackage{caption}

\usepackage{subcaption}

\newcommand{\be}{\begin{eqnarray}}
\newcommand{\ee}{\end{eqnarray}}
\newcommand{\beq}{\begin{equation}}
\newcommand{\eeq}{\end{equation}}
\newcommand{\bemul}{\begin{multline}}
\newcommand{\eemul}{\end{multline}}

\journal{Astroparticle Physics}

\begin{document}

\begin{frontmatter}



\title{Neutrinos from super-Eddington Seyfert galaxies}


\author[second1,second2]{Lucas M. Pasquevich}
\author[second1,second2]{Gustavo E. Romero}
\affiliation[second1]{organization={
Instituto Argentino de Radioastronomía (CCT La Plata, CONICET; CICPBA; UNLP)},
            addressline={C.C.5}, 
            city={Villa Elisa},
            postcode={1894}, 
            state={Provincia de Buenos Aires},
            country={Argentina}}   
\affiliation[second2]{organization={
Facultad de Ciencias Astronómicas y Geofísicas, Universidad Nacional de La Plata,  La Plata, Argentina},
            addressline={Paseo del Bosque S/N}, 
            city={La Plata},
            postcode={1900}, 
            state={Provincia de Buenos Aires},
            country={Argentina}}    
\author[first,first2]{Mat\'ias M. Reynoso}
\affiliation[first]{organization={Instituto de Investigaciones Físicas de Mar del Plata (IFIMAR – CONICET), and Departamento de Física, Facultad de Ciencias Exactas y Naturales, Universidad Nacional de Mar del Plata},
            addressline={Funes 3350}, 
            city={Mar del Plata},
            postcode={7600}, 
            state={Provincia de Buenos Aires},
            country={Argentina}}

\affiliation[first2]{organization={Departamento de Física, Facultad de Ciencias Exactas y Naturales, Universidad Nacional de Mar del Plata},
            addressline={Funes 3350}, 
            city={Mar del Plata},
            postcode={7600}, 
            state={Provincia de Buenos Aires},
            country={Argentina}}

\begin{abstract}

Multimessenger observations suggest that Seyfert galaxies are promising sources of high-energy neutrinos, but their dense inner environments can strongly suppress the emerging very-high-energy gamma-ray emission. Active galactic nuclei (AGN) undergoing intense accretion episodes can enter a super-Eddington state, in which the accretion flow becomes geometrically and optically thick within a critical radius and develops strong magnetic fields in its innermost region. At the same time, large amounts of matter are expelled from the disk surface in the form of powerful, radiation-driven winds. In this work, we explore a scenario in which the cores of super-Eddington AGN provide suitable conditions for the acceleration of relativistic particles, including hadrons, via magnetic reconnection in a magnetically confined region close to the supermassive black hole. The accelerated hadronic component interacts with the intense photon field of the disk, leading to a copious neutrino flux peaking at 10-100 TeV that may be detectable with current observatories such as IceCube and KM3NeT, while the surrounding outflow efficiently absorbs the accompanying gamma-ray emission from the inner core. We also apply the model to the nearby super-Eddington Seyfert 1 NGC 7469 as a representative case with two reported neutrino events. In this framework, super-Eddington AGN, in particular Seyfert galaxies undergoing transient intense accretion episodes, emerge as plausible hidden neutrino sources, offering a natural explanation for the coexistence of efficient neutrino production and a strongly attenuated gamma-ray counterpart.

\end{abstract}



\begin{keyword}
Neutrinos \sep Super-Eddington accretion \sep active galactic nuclei \sep multi-messenger astronomy 



\end{keyword}

\end{frontmatter}




\section{Introduction} \label{sec:intro}

Multiwavelength observations are essential for constraining the physical conditions and emission mechanisms of astrophysical sources. In recent decades, this approach has naturally evolved into multimessenger astronomy, driven in particular by the development of high-energy neutrino astrophysics and by the growing capability of neutrino observatories to identify potential Galactic and extragalactic sources. Unlike photons, which can be efficiently absorbed and reprocessed in dense, highly radiative environments, neutrinos escape essentially unaffected, thus providing a direct probe of hadronic activity deep inside compact systems.

High-energy neutrinos are produced through interactions of relativistic hadrons with either ambient matter ($pp$) or radiation fields ($p\gamma$), generating charged mesons that subsequently decay into neutrinos. These mechanisms are accompanied by electromagnetic emission, for example through the decay of neutral pions, with luminosities that can be comparable to the neutrino emission. In contrast, purely leptonic scenarios, in which relativistic electrons interact with ambient fields via synchrotron radiation, inverse Compton (IC) scattering, and relativistic Bremsstrahlung, can account for electromagnetic emission but not for efficient neutrino production. Therefore, the detection of an astrophysical neutrino signal provides direct evidence for hadronic processes and offers a powerful way to discriminate among leptonic, hadronic, and hybrid emission scenarios.

Active galactic nuclei (AGN) are among the most compelling candidate sources of TeV–PeV neutrinos. In the innermost regions around the supermassive black hole (SMBH), non-thermal particles might be accelerated in the ambient plasma, while the local radiation and matter fields provide suitable targets for efficient hadronic interactions. The first association of the $\sim 290$ TeV event IceCube-170922A with the flaring blazar TXS 0506+056 established AGN as relevant multimessenger neutrino sources \citep{TXS_0506_Aartsen_2020}. Beyond blazars, IceCube has reported steady neutrino emission from the nearby Seyfert galaxy NGC 1068, with a significance of $4.2\sigma$ \citep{IC_NGC1068_2022}. A remarkable feature of NGC 1068 is the pronounced mismatch between its neutrino and gamma-ray signal. The inferred neutrino luminosity, $L_\nu \sim 3\times10^{42}\,{\rm erg\,s^{-1}}$ in the $\sim {\rm TeV}$–$10\,{\rm TeV}$ range, exceeds the observed $0.1$–$100$ GeV gamma-ray luminosity, $L_\gamma \sim 10^{41}\,{\rm erg\, s^{-1}}$, while very-high-energy gamma rays above $\sim 200$ GeV remain undetected \citep{NGC1068_MAGIC_col2019,IC_NGC1068_2022}. This neutrino–gamma-ray gap suggests that the dominant neutrino production region is embedded in an environment opaque to high-energy photons, such that gamma rays are absorbed and reprocessed while neutrinos escape freely. Recent analyses have also reported neutrino excesses from other Seyfert galaxies, including NGC~4151 in the Northern Hemisphere and the Circinus galaxy in the Southern Hemisphere, although their interpretation as neutrino counterparts remains less secure \citep{IC_Seyferts_2026}.

A number of theoretical studies have proposed that obscured AGN cores can operate as efficient hidden hadronic accelerators, producing neutrinos while suppressing the emerging gamma-ray flux through interactions with dense radiation and matter fields \citep{HidenCoresAGN_Murase_2020,NGC1068nu_Inoue_2020,SeyfertNu_Kheirandish_2021,HiddenHearts_Murase_2022,Seyferts_Murase_2024,NGC1068Nu_Reis_2025,SeyfertNu_Wang_2026}. In this context, the accretion regime becomes a key ingredient. While most current research has focused on sub-Eddington AGN, SMBHs accreting above the Eddington rate are expected to develop optically thick inner flows and powerful radiatively driven winds. Such conditions naturally provide both intense target photon fields for $p\gamma$ interactions and large opacities for gamma rays, making super-Eddington AGN especially attractive as hidden neutrino sources and as a possible explanation for the coexistence of bright neutrino emission with a strongly reprocessed gamma-ray counterpart.

Accretion regimes are commonly characterized by the mass accretion rate relative to the Eddington limit,
\begin{equation}
\dot M_{\rm Edd}=\frac{L_{\rm Edd}}{\eta c^{2}},
\end{equation}
where $\eta\simeq 0.1$ is the accretion efficiency and
\begin{equation}
L_{\rm Edd}=\frac{4\pi G M_{\rm BH} m_{\rm p} c}{\sigma_{\rm T}}
\end{equation}
is the Eddington luminosity. Here $\sigma_{\rm T}$ is the Thomson cross section, $c$ is the speed of light, $m_{\rm p}$ is the proton mass, $G$ is the gravitational constant, and $M_{\rm BH}$ is the black hole mass. Super-Eddington accretion ($\dot M_{\rm input}>\dot M_{\rm Edd}$) is expected to be especially relevant for relatively low-mass SMBHs ($M_{\rm BH}\lesssim10^{7}\,M_\odot$), and such systems have been explored both observationally and theoretically in the context of supercritical AGN and quasars \citep{SuperEDD_Quasars_Liu_2019,SMBHEdd_Jiang_2019,SuperEdd_Quasars_Berdina_2021,SuperEddSMBH_Asahina_2022,SMBHEdd_Sotomayor_2022,Abaroa&Romero2024smbh}. Seyfert galaxies, and particularly narrow-line Seyfert 1 (NLS1), are often discussed as systems accreting at high Eddington ratios \citep{NLS1Revisited_Liu_2016,NLS1Edd_Collin_2004,fukue2004,SeyfertEdd_Jin_2009,IRAS_04416_Tortosa_2022,EddNLS1_Jin_2023,NLS1_Jiang_2025}. In these systems, strong absorption and complex radiative transfer can introduce substantial uncertainty in the accretion rates inferred for individual sources, including several well-studied Seyfert galaxies proposed as potential super-Eddington accretors.

Sub-Eddington AGN in the local Universe may also experience transient episodes of intense accretion triggered by tidal disruption events (TDEs), in which a star is disrupted by the tidal field of the central SMBH. During these episodes, the fallback rate can temporarily exceed the Eddington limit and may even reach hypercritical values, $\dot M\approx 10^3$--$10^4\dot M_{\rm Edd}$, depending on the properties of both the disrupted star and the SMBH \citep{Review_TDE_Gezari_2021,TDE_Kaur_2025,TDE_Price_2024}. The associated emission can persist from months to several years, whereas the characteristic fallback timescale for a solar-type star disrupted by an SMBH of mass $M_{\rm SMBH}\approx 10^6M_\odot$ is $\sim 40$ days \citep{Review_TDE_Gezari_2021}. For more massive stars and SMBHs, the super-Eddington phase can be substantially longer. For example, in the disruption of an O5 star with $M_\star=37\,M_\odot$ and $R_\star=11\,R_\odot$ by an SMBH of mass $M_{\rm SMBH}\approx 10^7\,M_\odot$, the accretion rate can peak above $\dot M>10^2\dot M_{\rm Edd}$ and remain super-Eddington for $t_{\rm Edd}\approx 7$ yr \citep{Abaroa&Romero2024smbh}. TDEs therefore provide a natural channel for producing transient super-Eddington accretion phases in SMBHs with $M_{\rm SMBH}\lesssim10^{7.5}\,M_\odot$.

In recent years, neutrino production in super-Eddington X-ray binaries has been actively explored, particularly in the context of ultraluminous X-ray sources \citep{Asthana_2023,Mushtukov2025,ULXMlti_Peretti_2025,ULXNeutrinos_Pasqa_2026}. In addition, recent theoretical work has studied neutrino production in the broad-line region of supercritical AGN, where the emission is generated through $pp$ interactions far from the central core \citep{neutrinos_Sotomayor_Romero2025}. Most AGN neutrino models, however, have been developed for sub-Eddington systems, usually involving geometrically thin, optically thick disks with a hot corona in the innermost region. Here we explore a different scenario, in which relativistic particles are confined in the magnetized innermost region of a supercritical accretion flow around an SMBH, where magnetic reconnection can operate as an efficient acceleration mechanism. The accelerated hadrons interact predominantly with the ambient photon field from the supercritical disk via $p\gamma$ processes, producing neutrinos, while the accompanying high-energy photons are strongly attenuated by the surrounding wind. In this framework, super-accreting AGN can behave as efficient sources of very-high-energy neutrinos even in the absence of relativistic jets.

This paper is organized as follows. In Sect. \ref{sec:model} we describe the structure of the supercritical disk and its associated wind. In Sect. \ref{sec:cool_distro} we introduce the injection of relativistic particles, compute their energy losses, and obtain the particle distributions, along with the corresponding meson and secondary production. In Sect. \ref{sec:radiation and opacity} we calculate the opacity and the resulting spectral energy distributions. In Sect. \ref{sec:neutrino_flux} we derive the neutrino emission and the corresponding flux at Earth, as well as its detectability with current and next-generation neutrino observatories such as IceCube, IceCube-Gen2, and KM3NeT. In Sect. \ref{sec:appl} we apply the model to the Seyfert 1 source NGC 7469. Finally, in Sect. \ref{sec:discussion} we discuss the main implications of the model and its observational signatures, and in Sect. \ref{sec:conclusion} we summarize our conclusions.

\section{Basics of the model} \label{sec:model}
We consider a SMBH of mass $M_{\rm BH}=10^{7}\,M_\odot$ accreting at a supercritical rate, $\dot m=\dot M_{\rm input}/\dot M_{\rm Edd}>1$. In order to explore the conditions under which such systems can operate as hidden neutrino sources, we describe below the structure of the supercritical disk, the associated wind, and the compact magnetized region where non-thermal particles are accelerated.

\subsection{Supercritical disk structure}

In the super-Eddington regime, the accretion flow develops a large optical depth and high radial inflow velocities. In this regime, photon trapping becomes important because the advection timescale becomes shorter than the photon diffusion timescale \citep{ohsuga2005,Kitakietal2021}. As a consequence, a significant fraction of the dissipated energy is advected inward with the flow and eventually accreted into the BH, thereby reducing both the emergent luminosity and the radiative efficiency. We describe this effect through the standard partition $Q_{\rm adv}=Q_{\rm vis}-Q_{\rm rad}=f\,Q_{\rm vis}$, where $Q_{\rm vis}$ is the viscous heating rate, $Q_{\rm rad}$ is the radiative cooling rate, and $f$ is the advection parameter \citep{narayan1994}.

The characteristic radius inside which advection and wind mass loss become relevant can be written as \citep{fukue2004}
\begin{equation}
r_{\rm crit}=\frac{9\sqrt{3}\sigma_{\rm T}\dot M_{\rm input}}{16\pi m_{\rm p}c}
\approx 40\,\dot m\,r_{\rm g},
\end{equation}
where $r_{\rm g}=GM_{\rm BH}/c^{2}$ is the gravitational radius. For $r>r_{\rm crit}$, the disk remains optically thick and geometrically thin, as in the standard Shakura-Sunyaev solution \citep{shakura1973}. For $r<r_{\rm crit}$, the flow becomes geometrically and optically thick, and radiation-driven winds are launched from the disk surface. In this region, the outflow regulates the accretion rate according to
\begin{equation}\label{eq:mass-accretion}
\dot M(r)=\dot M_{\rm input}\left(\frac{r}{r_{\rm crit}}\right)^{s+1/2},
\end{equation}
where the parameter $s$ accounts for the effect of mass loss through winds.

We follow \citet{fukue2004} and \citet{akizuki2006} for the description of the supercritical disk, which is based on the conservation laws of mass, radial momentum, angular momentum, and energy. In this formulation, the disk structure is determined by the BH mass ($M_{\rm BH}$), the accretion rate ($\dot m$), and a set of global parameters: the viscosity parameter ($\alpha$), the advection parameter ($f$), the adiabatic index ($\gamma$), the disk magnetization ($\beta_{\rm disk}$), and the wind ejection parameter $s$, which we set to $s=0.5$ when winds are present.

In the magnetized supercritical solution of \citet{akizuki2006}, the inclusion of an induction equation allows one to define a magnetization radius $r_{\rm mag}$ within the advection-dominated region ($r_{\rm mag}<r_{\rm crit}$). This radius depends on the magnetic flux accumulated near the BH \citep{McKinney_etal_2012}:
\begin{equation}
\begin{array}{l}
    r_{\rm mag} = r_{\rm g} \left[1.2\times 10^4 \left( \dfrac{3}{4} + \dfrac{n}{2} \right) \right]^{\frac{4}{3+2n}} \ \times \\ \\
    \times \ \ \epsilon_{-1}^{\frac{2}{3+2n}} m_8^{-\frac{6}{3+2n}} \dot{m}_{\rm H}^{-\frac{2}{3+2n}}\left( \dfrac{\Phi}{0.1\,{\rm pc^2\,G}} \right)^{\frac{4}{3+2n}},
\end{array}
\end{equation}
where $n=1$ characterizes the radial dependence of the accretion rate, $\dot M(r)\propto r^{n}$ (see eq. \eqref{eq:mass-accretion}), $\dot m_{\rm H}$ is the normalized accretion rate at the inner disk edge, $\epsilon_{-1}=\epsilon/0.1$ with $\epsilon\sim0.01$ relating the radial inflow speed to the free-fall velocity, $m_{8}\equiv M_{\rm BH}/(10^{8}M_\odot)$, and $\Phi$ is the magnetic flux near the BH.

To estimate $\Phi$, we adopt the magnetically arrested disk (MAD) flux as a reference scale \citep{RLRQ_AGN_Chamani2021}:
\begin{align}
\Phi_{\rm BH,MAD}\simeq 2.4\times 10^{34}
\left(\frac{\eta}{0.4}\right)^{-1/2}
\left(\frac{M_{\rm BH}}{10^9\,M_\odot}\right)
\nonumber\\
\times
\left(\frac{L_{\rm acc}}{1.26\times 10^{47}\,{\rm erg\,s^{-1}}}\right)^{1/2}
\ {\rm G\,cm^2},
\end{align}
where in our super-Eddington model $\eta=0.1$, $L_{\rm acc}=L_{\rm Edd}(1+\ln\dot m)$ \citep{King_rev_2023NewAR..9601672K}, and $M_{\rm BH}=10^{7}\,M_\odot$, we obtain $\Phi_{\rm MAD}=1.6\times10^{-5}\,{\rm G\,pc^{2}}$. The launching of relativistic jets requires an accretion flow able to accumulate a large-scale magnetic flux close to the MAD state \citep{Tchekhovsky_JetsMAD_2011,McKinney_etal_2012,RLRQ_AGN_Chamani2021}. The vast majority of Seyfert galaxies are classified as radio-quiet or remain undetected in the radio band, and only a small fraction of NLS1 galaxies are radio-loud or display weak jet activity \citep{Komosa_NLS1_RQ_2015,NLS1Nature_Chen_2018,Berton_NLS1RQ_2020}. Their magnetic fluxes are expected to lie well below the MAD value, which is associated with radio-loud systems. Because radio-loud and radio-quiet AGN can differ in radio output by typical factors of $10^3-10^4$ \citep{Panessa_RQAGNS_2019}, and synchrotron emission scales as $B^{2}$, this suggests a substantially weaker magnetic field in radio-quiet sources. Since the magnetic flux scales linearly with $B$, we suppose
\begin{equation}
\Phi \simeq 0.01\,\Phi_{\rm MAD}.
\end{equation}
We stress that this choice is not intended as a universal prescription, but as a fiducial sub-MAD normalization motivated by the radio-quiet nature of the Seyfert nuclei considered here. It should therefore be regarded as a conservative estimate for a radio-quiet Seyfert core. Larger magnetic fluxes would expand the magnetically dominated region of the disk, leading to a larger $r_{\rm mag}$, whereas in our reference scenario this region extends only over a few gravitational radii.

In the supercritical region, the disk scale height grows approximately linearly with radius, $H\propto r$. The semi-opening angle $\delta$ is then parameterized through the coefficient $c_{3}(\alpha,\beta_{\rm disk},s)$ as \citep{akizuki2006}
\begin{equation}
\tan\delta=\sqrt{c_{3}},
\end{equation}
where $c_{3}$ depends on the global disk parameters.

As discussed above, very high magnetization levels are not expected in these systems, and extremely large accretion rates are unlikely to be sustained except in transient events such as TDEs. In this work, we therefore consider two representative configurations: a weakly super-Eddington, low-magnetization case with $\dot m=3$ (model $A$), and a moderately magnetized case with $\dot m=10$ (model $B$). The adopted disk parameters for both models are summarized in Table \ref{tab:model_disk_parameters}.

\begin{table}[t]
\centering
\small
\caption{Model parameters from the supercritical disk.}
\label{tab:model_disk_parameters}
\begin{tabularx}{\columnwidth}{@{}>{\raggedright\arraybackslash}X c c c@{}}
\hline
\textbf{Parameter (symbol)} & Model $A$ & Model $B$ & Units \\
\hline
Mass accretion rate ($\dot m$)   & 3  & 10   & [$\dot M_{\rm Edd}$] \\
Viscosity parameter ($\alpha$)   & 0.1  & 0.01   &  \\
Advection parameter ($f$)   & 0.5  & 0.5   &  \\
Magnetization parameter ($\beta_{\rm disk}$)   & 0.5  & 5   &  \\
Adiabatic index ($\gamma$)   & 1.333  & 1.333   &  \\
Wind ejection ($s$)   & 0.5  & 0.5   &  \\
Semi-opening angle ($\delta$)   & 17  & 30   & [°] \\
Innermost temperature ($T_{\rm disk}$)   & $4.6\times10^{5}$ & $4.3\times10^{5}$ & [$\mathrm{K}$] \\
Magnetic field ($B$)   & $7.7\times10^{3}$ & $2\times10^{4}$ & [$\mathrm{G}$] \\
\hline
\end{tabularx}
\end{table}

In the innermost disk region, the toroidal magnetic field and effective temperature are given by \citep{akizuki2006}
\begin{equation}
B_{\phi}=
\sqrt{
4\pi\Sigma G M_{\rm BH}
\frac{\beta_{\rm disk} c_{3}}{\sqrt{(1+\beta_{\rm disk})c_{3}}}
}\,
r^{s/2-1},
\end{equation}
\begin{equation}
T_{\rm eff}=2.181\times10^{7}
\left(\frac{c_{3}}{1+\beta}\right)^{1/8}
\left(\frac{M_{\rm BH}}{10\,M_\odot}\right)^{-1/4}
\left(\frac{r}{r_{\rm g}}\right)^{-1/2}
{\rm K},
\end{equation}
where $\Sigma$ is the surface density. From these expressions we obtain, at the base of the inner disk, $B_{\rm A}=7.7\times10^{3}\,{\rm G}$ and $T_{{\rm disk},A}=4.6\times10^{5}\,{\rm K}$ for model $A$, and $B_{\rm B}=2\times10^{4}\,{\rm G}$ and $T_{{\rm disk},B}=4.3\times10^{5}\,{\rm K}$ for model $B$.

\subsection{Wind properties} \label{sec:wind_prop}

The wind is launched from the supercritical region of the disk, namely at radii $r<r_{\rm crit}$. Since the disk self-regulates its luminosity to values close to the Eddington limit, most of the accreted material is expected to be expelled from the disk surface in the form of radiation-driven winds, so that $\dot m_{\rm w}\approx\dot m_{\rm input}$. The terminal wind speed can be estimated as \citep{BHOutflowsKing_2010}
\begin{equation}
\beta_{\rm w}=\frac{1}{\sqrt{40\dot m_{\rm w}}}.
\end{equation}
Because $\beta_{\rm w}\ll 1$ and $\gamma_{\rm w}\approx 1$, the comoving and observed wind properties are essentially the same.

We assume that the outflow is approximately spherical, except for the narrow polar region where the wind is suppressed and a low-density funnel is formed around the $z$-axis of the system. In this picture, the wind constitutes the dense walls of the funnel, while the inner polar channel remains  transparent. The wind density is then obtained from the continuity equation as \citep{ObservationalBHWinds_Fukue2009,Abaroa-Romero2024RevMex,GalaxiesEDD_Abaroa_Romero_2024}
\begin{equation}\label{ec: density_wind}
\rho_{\rm w}=\frac{\dot M_{\rm w}}{4\pi R^{2}v_{\rm w}},
\end{equation}
where $R=\sqrt{r^{2}+z^{2}}$ is the distance to the BH.

Because the mass outflow rate is much larger than the Eddington accretion rate, the wind is optically thick and emits thermal radiation. The outflow remains opaque to its own radiation up to a height that defines the photosphere, $z_{\rm ph}$, above which photons can escape freely. The wind temperature at the photosphere can be estimated as \citep{ObservationalBHWinds_Fukue2009}
\begin{equation}\label{eq:T_field_wind}
\sigma_{\rm SB}T_{\rm w}^{4}=
\dot e\,
\frac{L_{\rm Edd}}{4\pi(r^{2}+z_{\rm ph}^{2})},
\end{equation}
where $\sigma_{\rm SB}$ is the Stefan--Boltzmann constant and $\dot e\sim0.1$ is the normalized luminosity in the comoving frame.


\begin{figure*}[h!]                            
\centering
  \centering
    \begin{subfigure}[t]{0.49\textwidth}
        \centering            \includegraphics[width=0.5\linewidth,trim= 130 30 120 20]{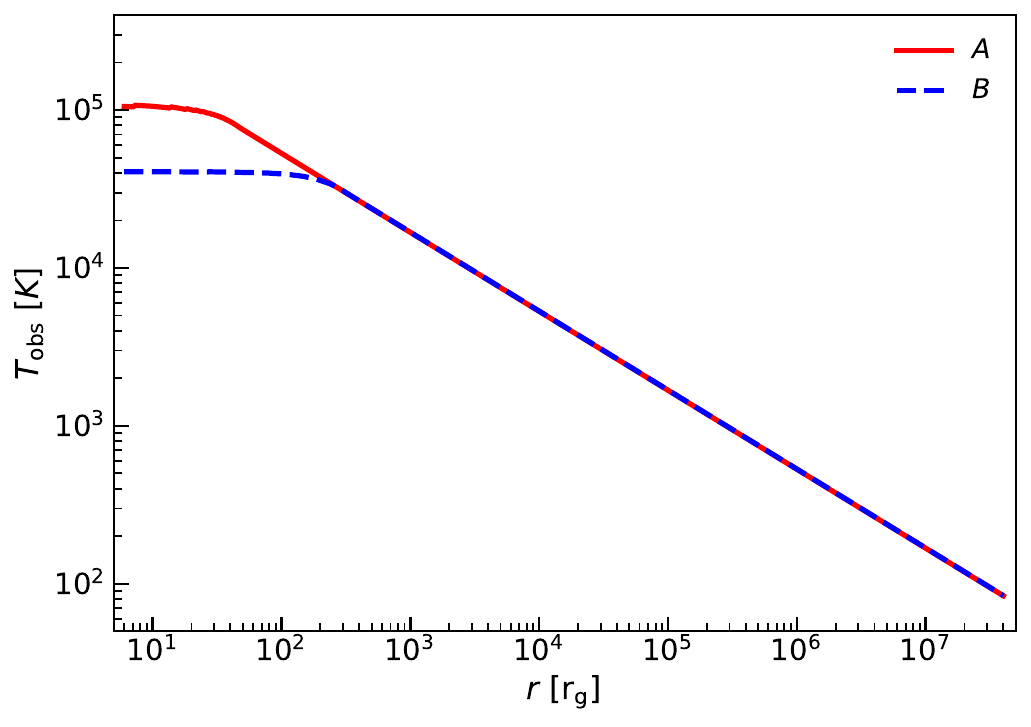} 
    \end{subfigure}
    \hfill
    \begin{subfigure}[t]{0.49\textwidth}
        \centering
    \includegraphics[width=0.5\linewidth,trim= 130 30 120 30]{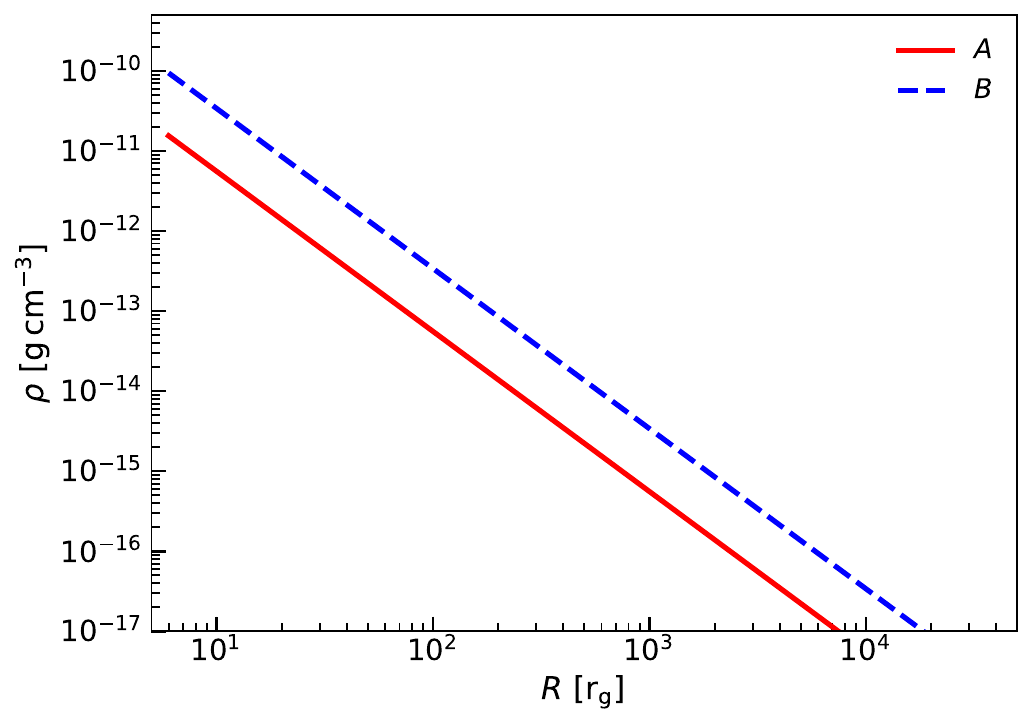} 
    \end{subfigure}
   
\caption{Wind temperature in the photosphere as a function of disk radius in $r_{\rm g}$ (left) and wind mass density as a function of radius from the center of the system in units of $r_{\rm g}$ (right).}

\label{fig:wind_prop}
\end{figure*}


Figure \ref{fig:wind_prop} shows the wind temperature at the photosphere and the wind mass density as a function of distance to the SMBH. Although the wind is launched outside the innermost magnetically confined region, and therefore is not expected to dominate the dynamics inside the acceleration zone, it plays a crucial radiative role in the system. In particular, the dense funnel walls and the wind photosphere provide an intense thermal photon field around the central core, with a characteristic temperature of $T_{\rm wind}\approx10^{5}\,{\rm K}$. Thus, in our scenario the wind affects the non-thermal processes mainly through its radiation field and through the absorption of high-energy photons, while the acceleration region itself is located inside the comparatively low-density funnel.

\subsection{Magnetically confined region}

In the vicinity of the SMBH, the magnetic field is expected to be highly turbulent, where small-scale magnetic reconnection events can develop through the interaction of accreting magnetic loops. Similar processes are observed in the solar corona \citep[e.g.][]{El2022A&A,Kimura2022ApJ,Karavola2025JCAP.}. Magnetic confinement mechanisms are expected on short spatial scales, comparable to the BH size \citep{CoronaCyg1_Romero2014,AGNNuSTAR_Fabian2015,AstroparticlesCoronae_Fang2024}. We adopt $z_{\rm acc}=10\,r_{\rm g}$ as a fiducial location for the acceleration region, corresponding to the innermost funnel where the radiation field is still intense and magnetic activity is expected to be strong, and we take $\Delta z_{\rm acc}=0.1\,z_{\rm acc}$ to represent a compact reconnection layer.

Seyfert spectra are often characterized, among other components, by thermal emission from the accretion disk. When this component is directly observed, it indicates that the line of sight crosses a relatively low-opacity channel and can intercept the innermost disk regions. In our geometrical picture, this requires the funnel to remain optically thin. The corresponding constraint on the Thomson optical depth is \citep{ObservationalBHWinds_Fukue2009,photosphere_Tomida2015}
\begin{equation}
\tau=\int_{0}^{\infty}\gamma_{\rm gas}\left(1-\frac{v_{\rm gas}}{c}\cos\vartheta \right)\kappa_{\rm co}\rho_{\rm co}\,dz<1,
\end{equation}
where $\rho_{\rm co}\approx\rho_{\rm gas}$ and $\kappa_{\rm co}\approx\kappa=\sigma_{\rm T}/m_{\rm p}$ are the density and opacity in the comoving frame, $\beta_{\rm gas}=v_{\rm gas}/c$ is the gas velocity in units of $c$, and $\gamma_{\rm gas}$ is the corresponding Lorentz factor. This condition yields an upper limit on the gas density of $n_{\rm gas}=1.2\times10^{10}\,{\rm cm^{-3}}$ for both models. This value should be regarded as a strict upper bound, since realistic sources are expected to have lower densities.

We adopt a one-zone description for the acceleration region, assuming uniform physical conditions within a layer of thickness $\Delta z_{\rm acc}=0.1\,z_{\rm acc}$. Protons and electrons are injected there with a power-law distribution and an exponential cutoff,
\begin{equation}
Q_{\rm e,p}(E_{\rm e,p})=Q^{0}_{\rm e,p}E_{\rm e,p}^{-\Gamma}
\exp\left(-\frac{E_{\rm e,p}}{E_{\rm e,p,{\rm max}}}\right),
\end{equation}
where $Q_{\rm e,p}(E_{\rm e,p})$ is the number of non-thermal particles injected per unit energy, solid angle, volume, and time. Here $Q^{0}_{\rm e,p}$ is a normalization constant, $\Gamma_{e,p}$ is the spectral index, and $E_{\rm e,p,{\rm max}}$ is the maximum particle energy, obtained by balancing the acceleration rate against the total energy-loss and escape rates. The normalization is fixed through the total injected non-thermal power, $L_{\rm e,p}$,
\begin{align}
L_{\rm e,p}
&=\int \mathrm{d}V \int \mathrm{d}\Omega \int \mathrm{d}E_{\rm e,p}\,
E_{\rm e,p}\,Q_{\rm e,p}(E_{\rm e,p})
\label{eq:Lp_def}\\
&=4\pi \Delta V \int \mathrm{d}E_{\rm e,p}\,
Q^0_{\rm e,p}\,E_{\rm e,p}^{-\Gamma+1}\exp\!\left(-\frac{E_{\rm e,p}}{E_{\rm e,p,\,max}}\right).
\end{align}
This relation determines $Q^{0}_{\rm e,p}$. In the last expression, the interaction volume is $\Delta V=\Delta z_{\rm acc}\,\pi r_{\rm f}^{2}$, where the transverse size of the region is given by $r_{\rm f}=\tan\delta\,z_{\rm acc}$, that is, the funnel radius at the height $z_{\rm acc}$.

We assume that $10\%$ of the magnetic power available in the reconnection region is converted into accelerated particles, so that $L_{\rm rel}=L_{\rm p}+L_{\rm e}=0.1\,L_{\rm mag}$ \citep{2011ApJ...738..115D}, with
\begin{equation}
L_{\rm mag}=\frac{B^2\, c\, z_{\rm acc}^2 \tan^2 \delta}{2}.
\label{eq:Lmag_def}
\end{equation}
We have estimated the area of the reconnection zone as $\sim\pi(z_{\rm acc}\tan\delta)^{2}$. With this prescription, we obtain a total non-thermal power of $L_{\rm A}\simeq5.3\times10^{43}\,{\rm erg\,s^{-1}}$ for model $A$ and $L_{\rm B}\simeq4.3\times10^{44}\,{\rm erg\,s^{-1}}$ for model $B$.
The partition of the non-thermal power between hadrons and leptons is not well constrained. The observed cosmic ray electron-to-proton ratio is $\sim10^{-2}$, indicating a proton-dominated population. Motivated by this fact, and by the strong radiative cooling experienced by electrons in the magnetized region, we adopt a fiducial power ratio $a=L_{\rm p}/L_{\rm e}=m_{\rm p}/m_{\rm e}\sim10^3$. The proton luminosity is therefore $L_{\rm p}=aL_{\rm rel}/(1+a)$, which depends only weakly on the value of $a$ as long as $a\gg1$.

A key quantity in this context is the gas magnetization, which we parameterize as \citep{2015SSRv..191..545K}
\begin{equation}
\sigma_{\rm gas}=\frac{B^{2}}{4\pi \,m_{\rm p}\, n_{i}\,c^2},
\end{equation}
where $B$ is the magnetic field and $n_i$ is the total particle number density in the confined region. The plasma is magnetically dominated when $\sigma_{\rm gas}>1$. The degree of magnetization is expected to affect the spectral slope of the accelerated particles, with softer spectra for lower magnetization and harder spectra for more strongly magnetized plasmas. For our low and moderate magnetization scenarios, we adopt $n_{i,{ A}}=4.4\times10^{8}\,{\rm cm^{-3}}$ and $\Gamma=2.5$ for model $A$, which yields $\sigma_{\mathrm{gas},A}=7$, and $n_{i,B}=2.1\times10^{9}\,{\rm cm^{-3}}$ and $\Gamma=2$ for model $B$, corresponding to $\sigma_{\mathrm{gas},B}=10$.

\section{Cooling rates and particle distribution} \label{sec:cool_distro}

\begin{figure*}[h!]                            
\centering
  \centering
    \begin{subfigure}[t]{0.49\textwidth}
        \centering            \includegraphics[width=0.5\linewidth,trim= 130 30 120 20]{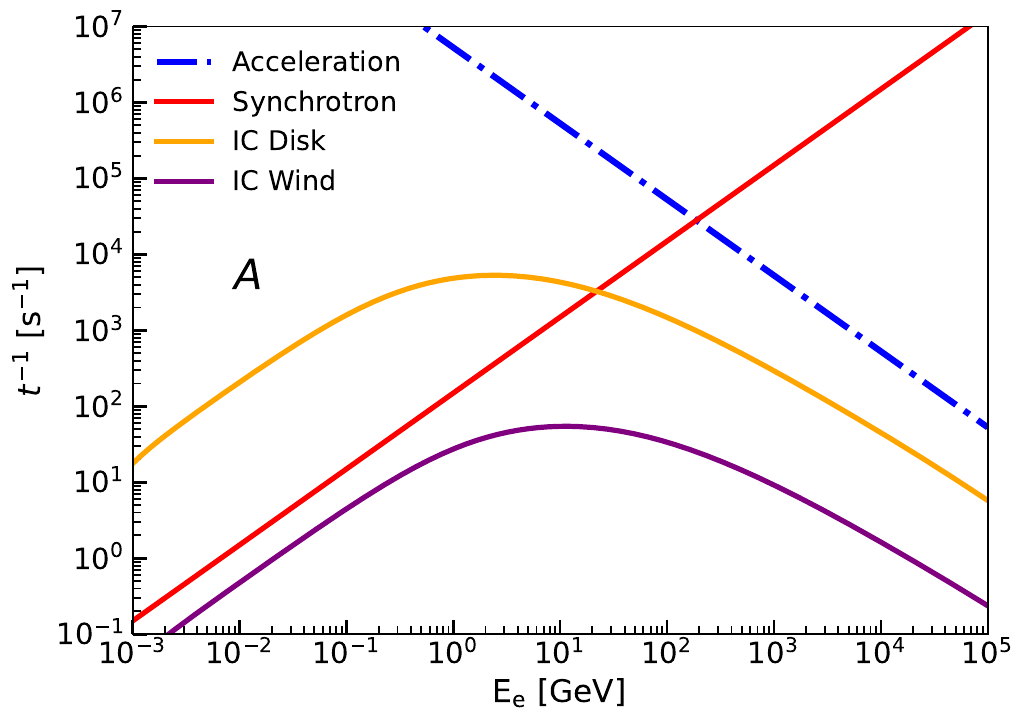} 
    \end{subfigure}
    \hfill
    \begin{subfigure}[t]{0.49\textwidth}
        \centering
    \includegraphics[width=0.5\linewidth,trim= 130 30 120 30]{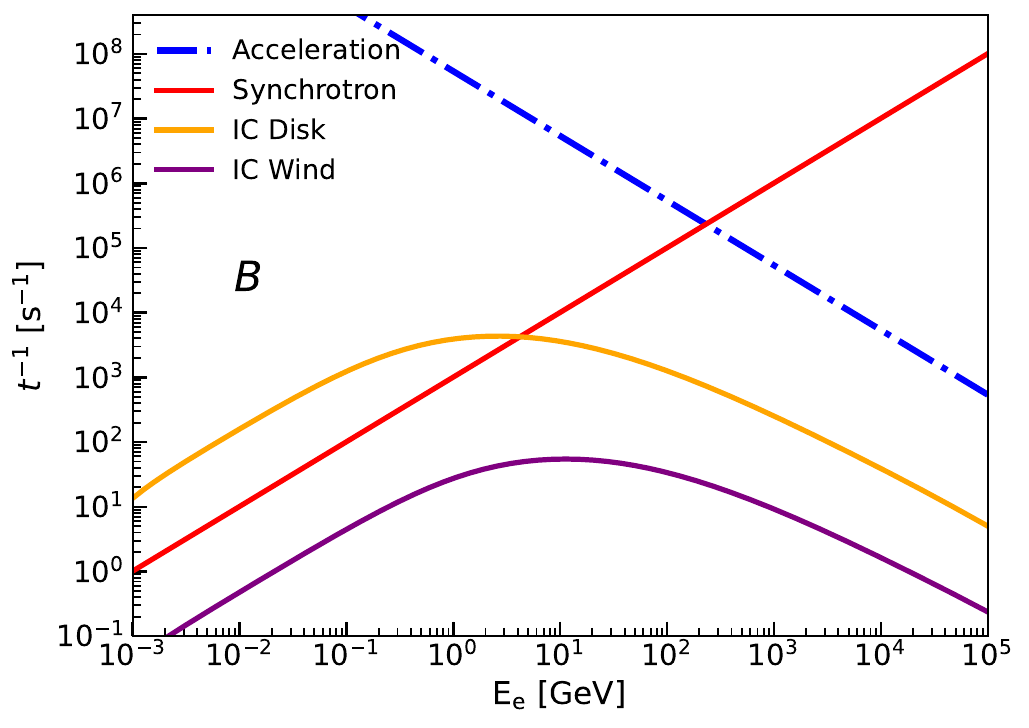} 
    \end{subfigure}
   
\caption{Acceleration, cooling, and diffusion rates for primary electrons. IC interactions with disk photons dominate at low energies, whereas synchrotron cooling becomes dominant above $\sim 20$ GeV in model A and above $\sim 3$ GeV in model B. In both cases, electrons reach maximum energies of $\sim 200$ GeV.}
 \label{fig:primary_rates_electron}
\end{figure*}

\begin{figure*}[h!]                            
\centering
  \centering
    \begin{subfigure}[t]{0.49\textwidth}
        \centering            \includegraphics[width=0.5\linewidth,trim= 130 30 120 20]{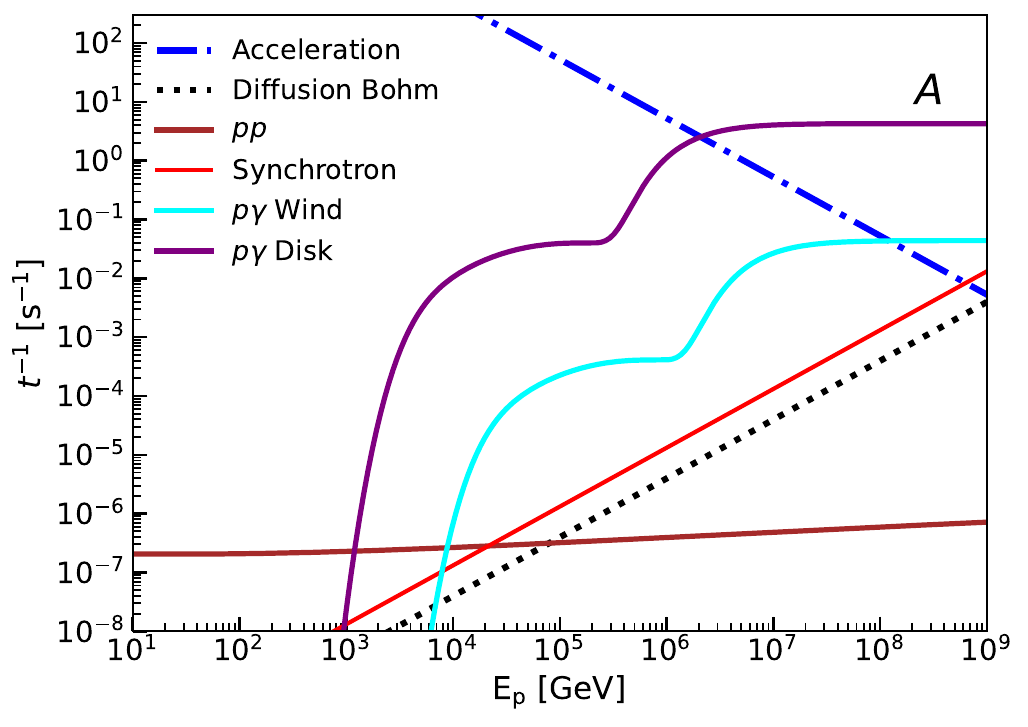} 
    \end{subfigure}
    \hfill
    \begin{subfigure}[t]{0.49\textwidth}
        \centering
    \includegraphics[width=0.5\linewidth,trim= 130 30 120 30]{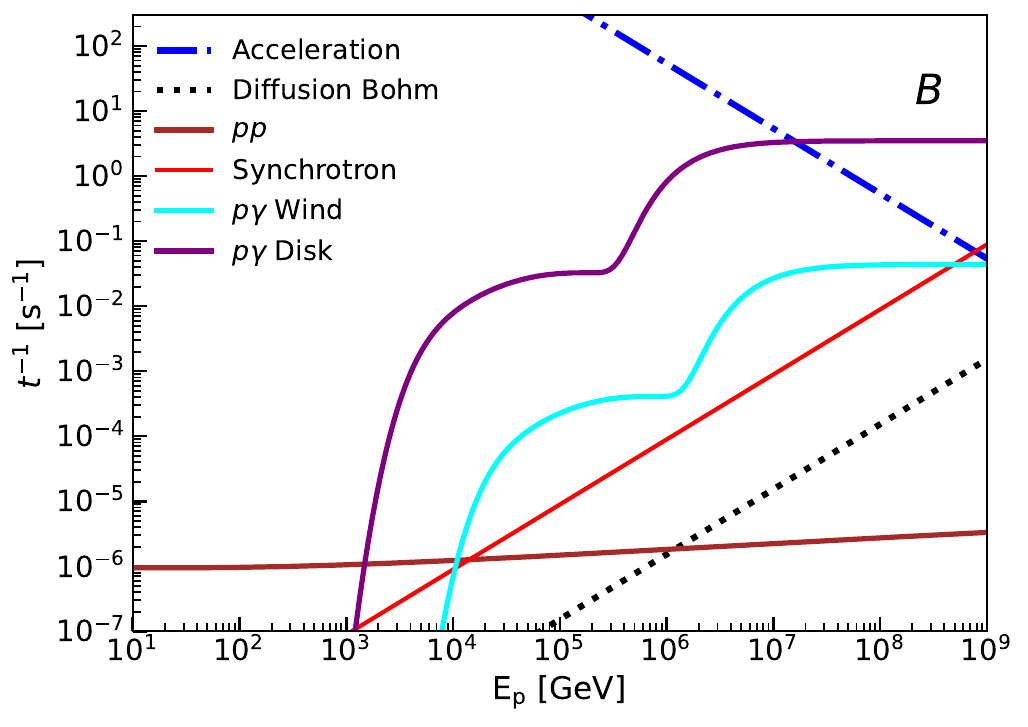} 
    \end{subfigure}
\caption{Acceleration, cooling, and diffusion rates for primary protons. At low energies, $pp$ interactions dominate the cooling, while above $\sim$TeV energies $p\gamma$ interactions become the main loss channel. The maximum proton energy reaches $\sim 1$ PeV in model A and $\sim 10$ PeV in model B.}

\label{fig:primary_rates_protons}
\end{figure*}

\begin{figure*}[h!]                            
\centering
  \centering
    \begin{subfigure}[t]{0.49\textwidth}
        \centering            \includegraphics[width=0.5\linewidth,trim= 130 30 120 20]{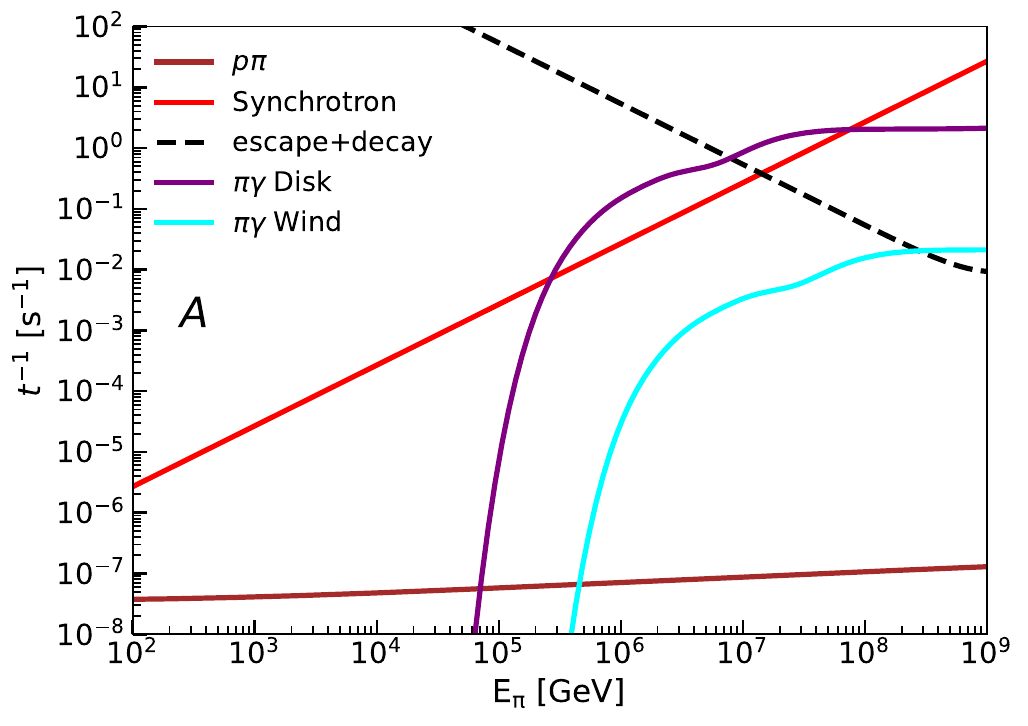} 
    \end{subfigure}
    \hfill
    \begin{subfigure}[t]{0.49\textwidth}
        \centering
    \includegraphics[width=0.5\linewidth,trim= 130 30 120 30]{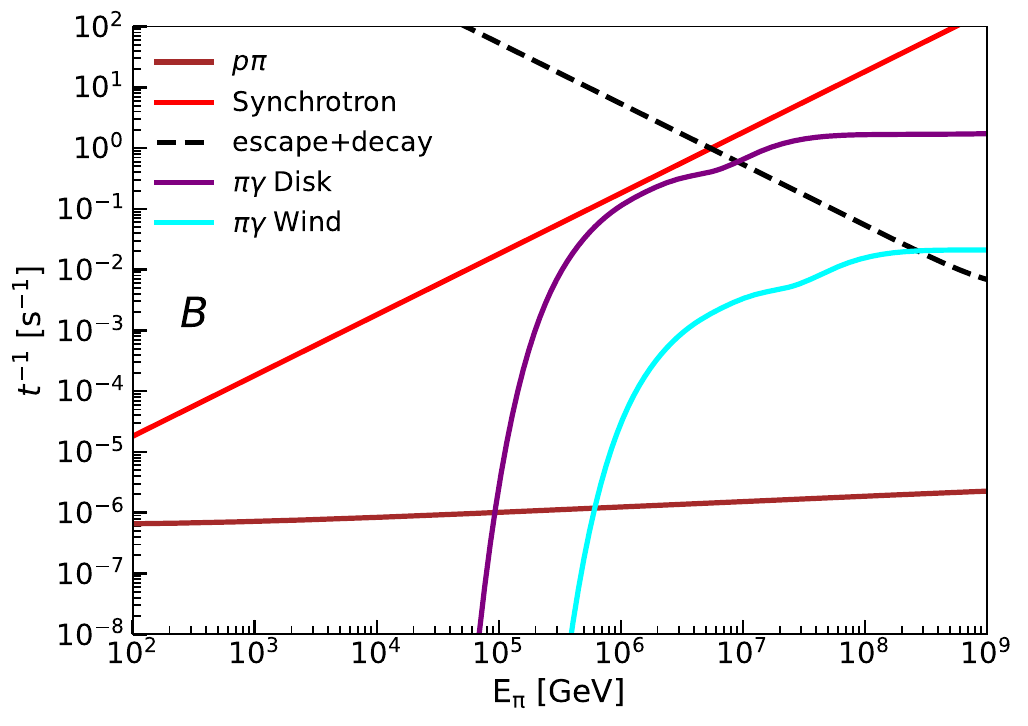} 
    \end{subfigure}
   
\caption{Cooling, escape, diffusion, and decay rates for charged pions. Synchrotron radiation provides the dominant cooling channel over most of the relevant energy range, while $\pi\gamma$ interactions become important above $\sim 2\times10^{5}$ GeV.}

\label{fig:pion_rates}
\end{figure*}

\begin{figure*}[h!]                            
\centering
  \centering
    \begin{subfigure}[t]{0.49\textwidth}
        \centering            \includegraphics[width=0.5\linewidth,trim= 130 30 120 20]{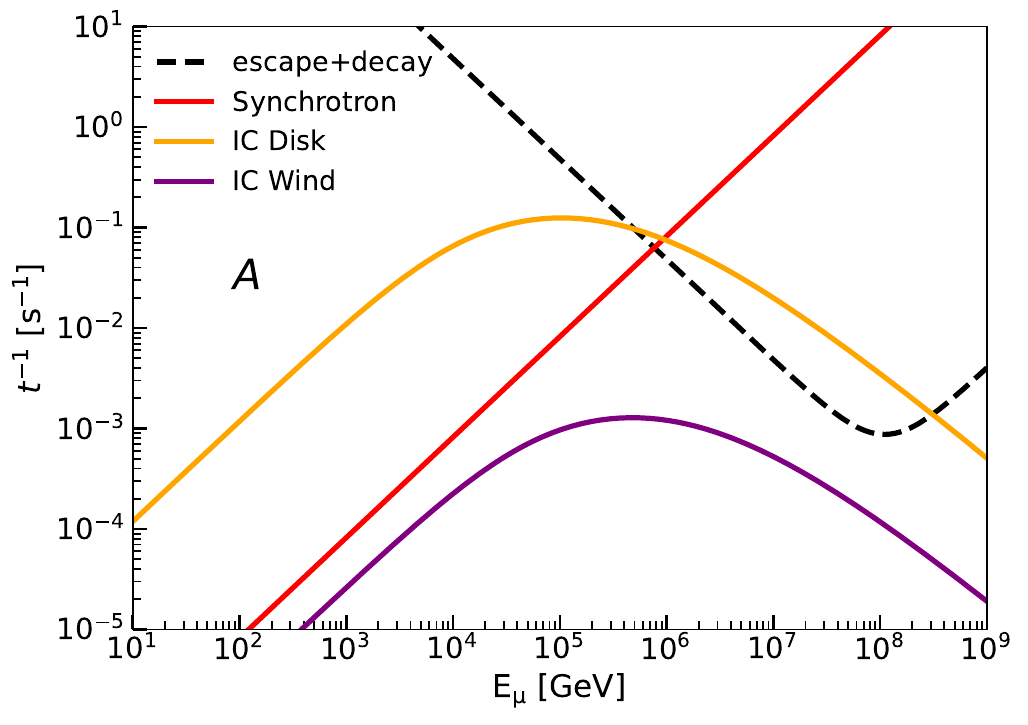} 
    \end{subfigure}
    \hfill
    \begin{subfigure}[t]{0.49\textwidth}
        \centering
    \includegraphics[width=0.5\linewidth,trim= 130 30 120 30]{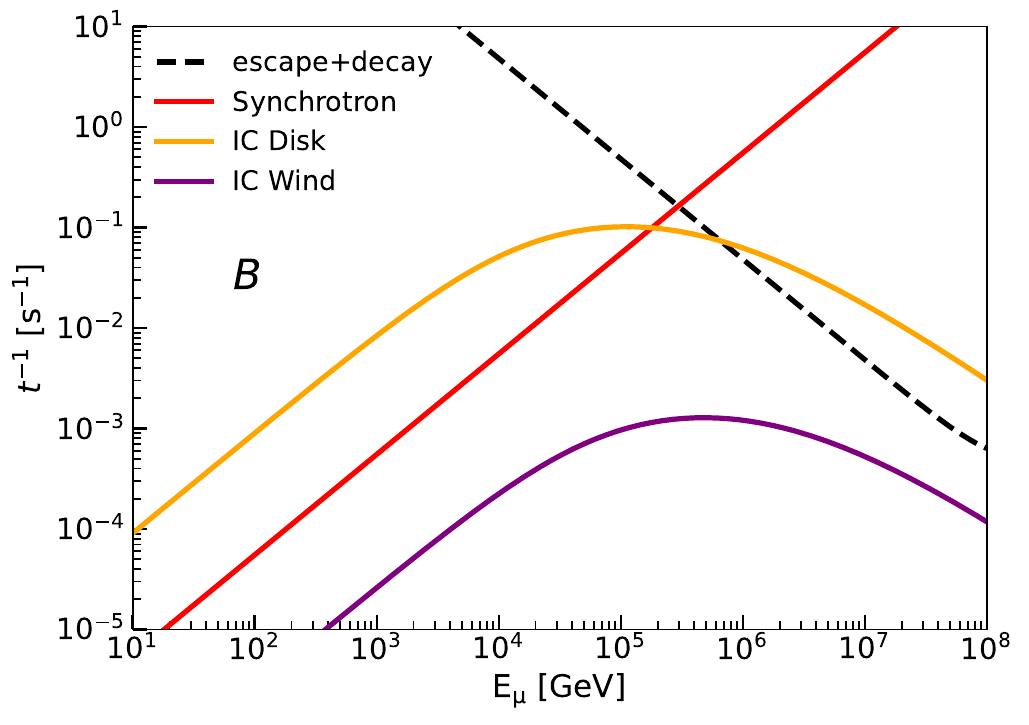} 
    \end{subfigure}
   
\caption{Cooling, escape, diffusion, and decay rates for muons.}

 \label{fig:muon_rates}
\end{figure*}

The acceleration rate of charged particles in the magnetized region is 
\begin{equation}
t_{\mathrm{acc}}^{-1} = \frac{\eta_{\rm rec} e c B}{E},
\label{equation_AccelerationRate}
\end{equation}
where $e$ is the elementary charge and $\eta_{\rm rec}$ is the acceleration efficiency. Since particle acceleration is assumed to be driven by magnetic reconnection, we adopt \citep{2011ApJ...738..115D,vieyro2012particle} 
\begin{equation}
\eta_{\rm rec}=0.1\,\frac{E v_{\rm rec}^{2}}{c\,D(E)\,eB},
\end{equation}
with $D(E)=Ec/(3eB)$ the Bohm diffusion coefficient. In the strongly magnetized conditions considered here, the reconnection speed is expected to be comparable to the Alfvén speed, $v_{\rm rec}\sim v_{\rm A}$, and in the relativistic regime $v_{\rm A}\sim c$, which yields $\eta_{\rm rec}\approx 0.3$.

We compute electron energy losses due to synchrotron radiation and IC. Bremsstrahlung losses are neglected because of the low gas density in the acceleration region. For protons, we include synchrotron losses, inelastic $pp$ collisions, and $p\gamma$ interactions with photons from the disk and the wind. These cooling processes are calculated using standard expressions given by \citet{begelman1990}, \citet{kelner2008}, and \citet{romero2008}.

In particular, the $p\gamma$ cooling rate for a target photon field $n_{\rm ph}(E)$ is defined as
\begin{equation} \label{eq:pgammarate}
t_{p\gamma}^{-1} = \frac{c}{2\gamma_p^2} \int_{E_{\rm th}/(2\gamma_p)}^{\infty} \mathrm{d}E\,\frac{n_{\rm ph}(E)}{E^{2}} \int_{E_{\rm th}}^{2\gamma_p E} \mathrm{d}E'\, \sigma_{p\gamma}(E')\,\kappa_{p\gamma}(E')\,E',
\end{equation}
where $\sigma_{p\gamma}$ is the total cross section and $\kappa_{p\gamma}$ is the corresponding inelasticity. This interaction includes two main channels: Bethe-Heitler pair production, $p+\gamma \rightarrow p+e^-+e^+$, and photomeson production. The latter has a threshold energy of $\sim 145\,{\rm MeV}$ in the proton rest frame and proceeds mainly through channels of the form
$p+\gamma\rightarrow p+a\pi^0+b(\pi^++\pi^-)$ and
$p+\gamma\rightarrow n+\pi^++a\pi^0+b(\pi^++\pi^-)$.
Neutrinos are mainly produced through the decay chains
$\pi^+\rightarrow\mu^++\nu_\mu\rightarrow e^++\nu_e+\bar{\nu}_\mu+\nu_\mu$
and
$\pi^-\rightarrow\mu^-+\bar{\nu}_\mu\rightarrow e^-+\bar{\nu}_e+\nu_\mu+\bar{\nu}_\mu$.
For the Bethe-Heitler process, the cross section $\sigma_{p\gamma}$ and inelasticity $\kappa_{p\gamma}$ are taken from \citet{Maximon1968}, while for photomeson production we follow \citet{romero2008}. We also include the production of charged kaons through $p\gamma$ interactions, with a threshold energy of order $\sim 1\,{\rm GeV}$. Although subdominant, this channel contributes at the highest energies through the decays $K^+\rightarrow\mu^+ +\nu_\mu$ and $K^-\rightarrow\mu^-+\bar{\nu}_\mu$.

We also compute the diffusion of charged particles in the turbulent region as
\begin{equation}
    t^{-1}_{\rm diff} = \frac{2D(E)}{L^2},
\end{equation}
where $D(E)$ is the diffusion coefficient and $L$ is the characteristic size of the region. As a reference, we consider the Bohm diffusion coefficient\footnote{An estimate of mirror diffusion following \citet{Lazarian_mirrors_2021}, evaluated for the same reconnection region, shows that the corresponding escape rate remains subdominant compared with the dominant cooling and the Bohm diffusion rate.}, $D_{\rm Bohm}(E) = E\,c / 3eB$.

Figures \ref{fig:primary_rates_electron} and \ref{fig:primary_rates_protons} show the acceleration and cooling rates of primary electrons and protons. For electrons, IC losses with disk photons dominate up to $\approx 20$ GeV in model $A$ and up to $\approx 3$ GeV in model $B$, while synchrotron cooling dominates at higher energies. For protons, inelastic $pp$ collisions dominate the losses up to $\sim$TeV energies, whereas above that range $p\gamma$ interactions with the disk photon field become the main cooling channel. Accordingly, electron acceleration is limited by synchrotron losses, whereas proton acceleration is limited mainly by $p\gamma$ interactions with disk photons. Diffusive escape, both in the Bohm and mirror diffusion regimes, remains subdominant over the relevant energy range. The resulting maximum energies are $E^{\rm max}_{\rm e}=200$ GeV and $E^{\rm max}_{\rm p}=1$ PeV for model $A$, while for model $B$ electrons reach the same maximum energy and protons attain $E^{\rm max}_{\rm p}=10$ PeV.

The cooling rates of charged pions and muons in the magnetically confined region are shown in Figs. \ref{fig:pion_rates} and \ref{fig:muon_rates}. Charged pions lose energy through $\pi p$ collisions, synchrotron radiation, and $\pi\gamma$ interactions with photons from the disk and the wind. The latter are computed with an expression analogous to eq. (\ref{eq:pgammarate}), using the cross sections and inelasticities discussed in \citet{Neutrinos_Rey_Deus_2023}. Muons, in turn, cool mainly through IC scattering with disk photons and through synchrotron radiation. For charged kaons, we include synchrotron losses and $K\gamma$ interactions, adopting the approximation $t^{-1}_{\pi^\pm\gamma}\approx t^{-1}_{K^\pm\gamma}$. We also compute the decay timescales of the unstable secondaries as
\begin{equation}
T_{\pi^\pm,\,\rm dec}=2.6\times10^{-8}\,\gamma_{\pi^\pm}\ {\rm s},
\end{equation}
\begin{equation}
T_{\mu^\pm, \,\rm dec}=2.2\times10^{-6}\,\gamma_{\mu^\pm}\ {\rm s},
\end{equation}
\begin{equation}
T_{K^\pm,\; \rm dec}=1.2\times10^{-8}\,\gamma_{K^\pm}\ {\rm s},
\end{equation}
where $\gamma_{\pi^\pm}$, $\gamma_{\mu^\pm}$, and $\gamma_{K^\pm}$ are the corresponding Lorentz factors.

\begin{figure*}[h!]                            
\centering
  \centering
    \begin{subfigure}[t]{0.48\textwidth}
        \centering            \includegraphics[width=0.5\linewidth,trim= 130 30 120 20]{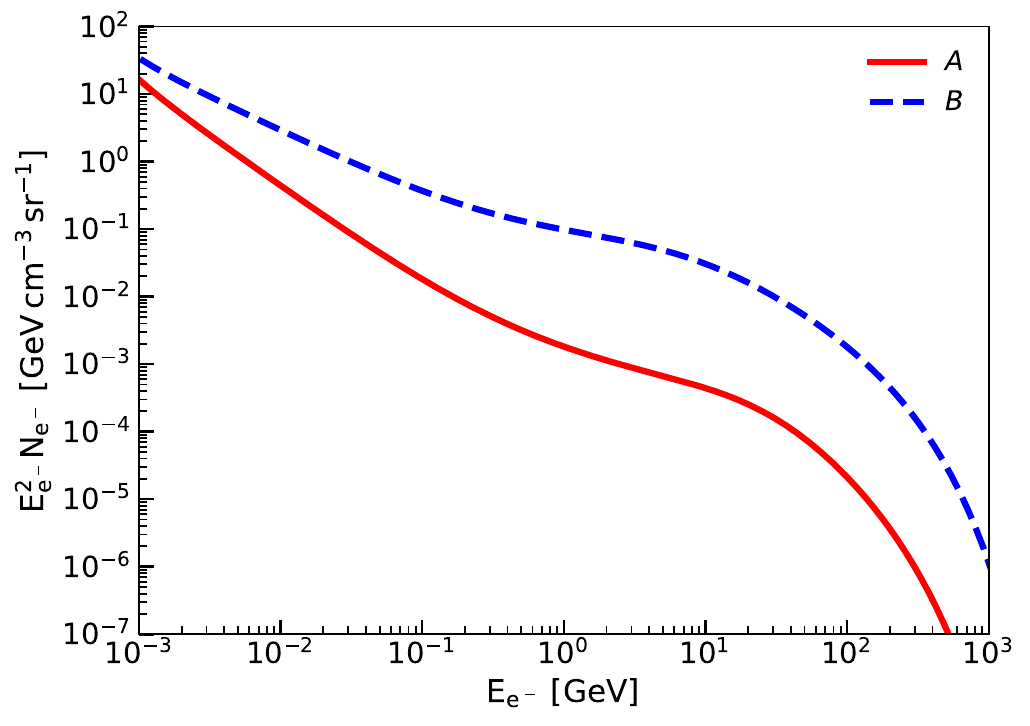} 
    \end{subfigure}
    \hfill
    \begin{subfigure}[t]{0.48\textwidth}
        \centering
    \includegraphics[width=0.5\linewidth,trim= 130 30 120 30]{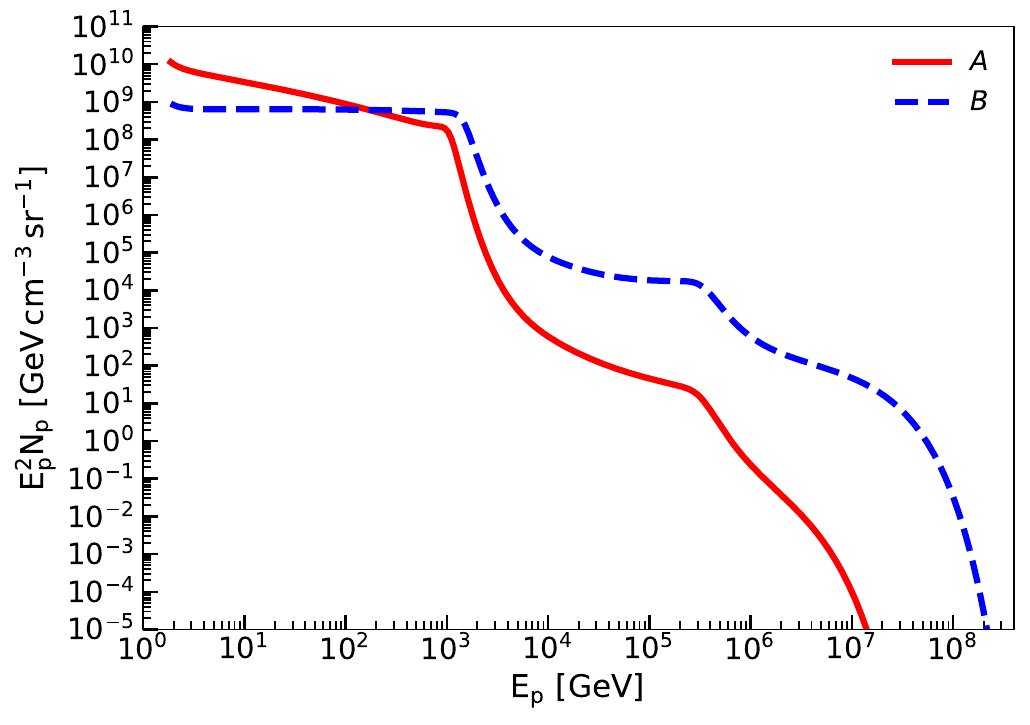} 
    \end{subfigure}
   
\caption{Steady-state distributions of primary electrons and protons.}
 \label{fig:primary_distro}
\end{figure*}

\begin{figure}
    \centering
    \includegraphics[width=0.9\linewidth]{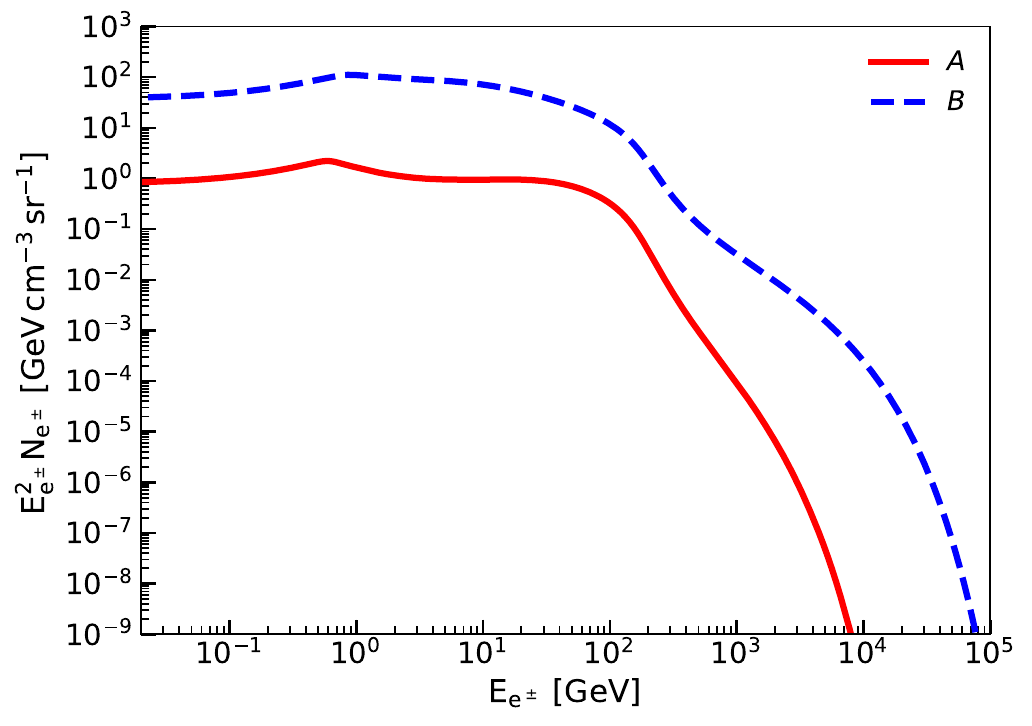}
\caption{Steady-state distribution of secondary pairs produced through the Bethe--Heitler process with disk photons as targets, for models A (solid red line) and B (dashed blue line).}
    \label{fig:distro_pares}
\end{figure}

\begin{figure}
    \centering
    \includegraphics[width=0.9\linewidth]{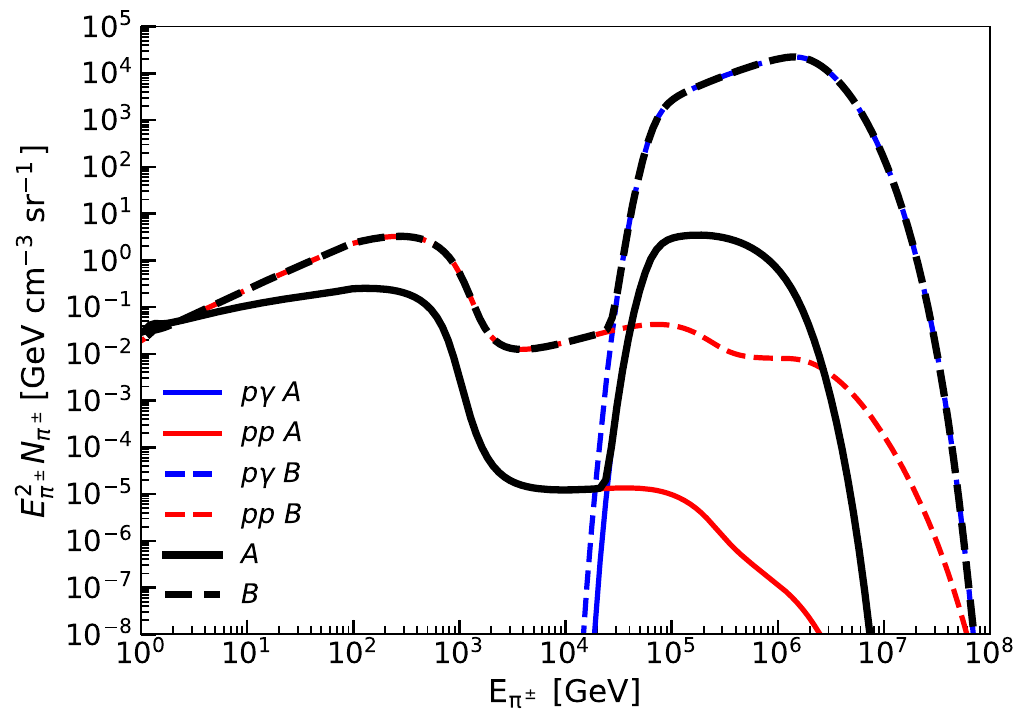}
\caption{Steady-state distributions of charged pions produced through $pp$ interactions (red lines) and $p\gamma$ interactions (blue lines), for model A (solid lines) and model B (dashed lines). The total distribution is shown in black in each case.}
    \label{fig:distro_piones}
\end{figure}

\begin{figure}
    \centering
    \includegraphics[width=0.9\linewidth]{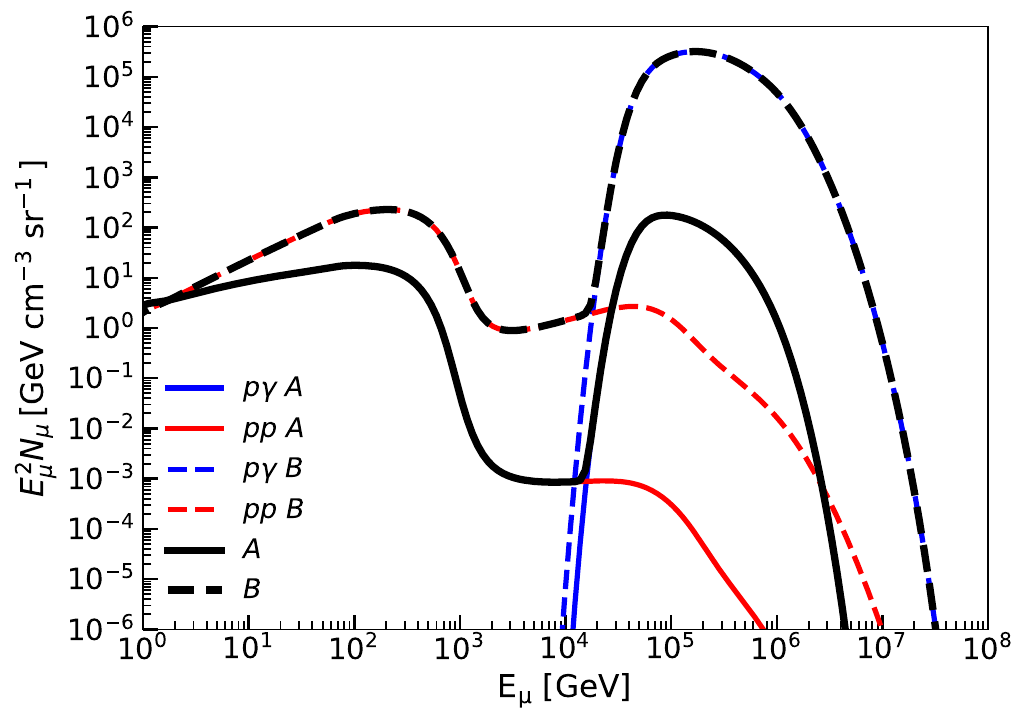}
\caption{Steady-state distributions of muons produced by pion decay from the $pp$ channel (red lines) and the $p\gamma$ channel (blue lines), for model A (solid lines) and model B (dashed lines). The total distribution is shown in black in each case.}
    \label{fig:distro_muones}
\end{figure}

\begin{figure}
    \centering
    \includegraphics[width=0.9\linewidth]{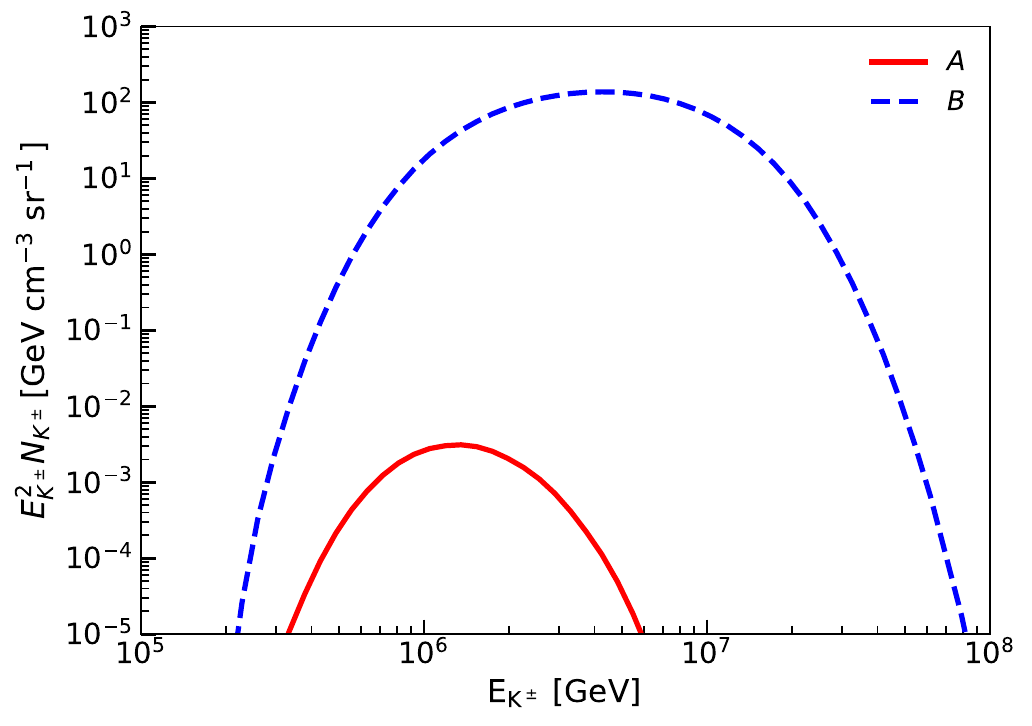}
\caption{Steady-state distribution of charged kaons produced through $p\gamma$ interactions with disk photons as targets, for model A (solid red line) and model B (dashed blue line).}
    \label{fig:distro_kaones}
\end{figure}

We obtain the steady-state particle distributions by solving a stationary one-zone transport equation for each species
$i\in\{p,e^-,e^\pm,\pi^\pm,\mu^\pm,K^\pm\}$ \citep{khangulyan2007}:
\begin{equation}
    \frac{\mathrm{d}\left[b_{i,\,\rm loss}(E_i) N_i(E_i)\right]}{\mathrm{d}E_i}+  \frac{N_i(E_i)}{T_{i,\rm esc}}= Q_i(E_i),\label{Eq_transport_i}  
\end{equation}
where the total energy loss reads
\begin{equation}
    b_{i,\, \rm loss}= -\left.\frac{\mathrm{d}E_i}{\mathrm{d}t}\right|_{\rm loss},
\end{equation}
including the contribution of all relevant processes. 
The resulting proton and electron distributions are shown in Fig. \ref{fig:primary_distro}. The corresponding distributions of secondary pairs, pions, muons, and kaons are shown in Figs. \ref{fig:distro_pares}-\ref{fig:distro_kaones}.

\begin{table}[t]
\centering
\small
\caption{Model parameters and values adopted in the text.}
\label{tab:model_parameters}
\begin{tabularx}{\columnwidth}{@{}>{\raggedright\arraybackslash}X c c c@{}}
\hline
\textbf{Parameter (symbol)} & Model $A$ & Model $B$ & Units \\
\hline
Wind temperature ($T_{\rm wind}$)   & $10^{5}$ & $10^{5}$   & $\mathrm{K}$ \\
Spectral index  ($\Gamma$)     & 2.5      & 2  &  \\
Magnetic power  ($L_{\rm mag}$)     & $1.9\times10^{43}$      & $4.3\times10^{44}$  & $\mathrm{erg\,s^{-1}}$ \\
Power in protons ($L_{\rm p}$)        & $1.9\times10^{42}$ & $4.3\times10^{43}$  & $\mathrm{erg\,s^{-1}}$ \\
Power in electrons ($L_{\rm e}$)      &  $1.9\times10^{39}$ & $4.3\times10^{40}$  & $\mathrm{erg\,s^{-1}}$ \\
Number density ($n_{i}$)           & $4.4\times10^{8}$ & $2.1\times10^{9}$   & $\mathrm{cm^{-3}}$ \\
Volume acc. region ($\Delta V$)                & $10^{38}$ & $3.7\times10^{38}$  & $\mathrm{cm^{3}}$ \\
Max. electron energy ($E^{\max}_{\rm e}$) & $2\times10^{11}$  & $2\times10^{11}$  & $\mathrm{eV}$ \\
Max. proton energy ($E^{\max}_{\rm p}$) & $8\times10^{14}$  & $10^{16}$  & $\mathrm{eV}$ \\
\hline
\end{tabularx}
\end{table}

\section{Opacity and SEDs} \label{sec:radiation and opacity}

We compute the non-thermal emission produced in the magnetically confined region by primary electrons and secondary pairs through synchrotron radiation and IC. We also include the gamma-ray emission generated by the decay of neutral pions produced in both $pp$ and $p\gamma$ interactions. For these radiative processes we adopt the standard expressions given by \citet{NT_Blumenthal_1970}, \citet{AyA_2000}, and \citet{romero2008}.


\begin{figure}
    \centering
    \includegraphics[width=0.9\linewidth]{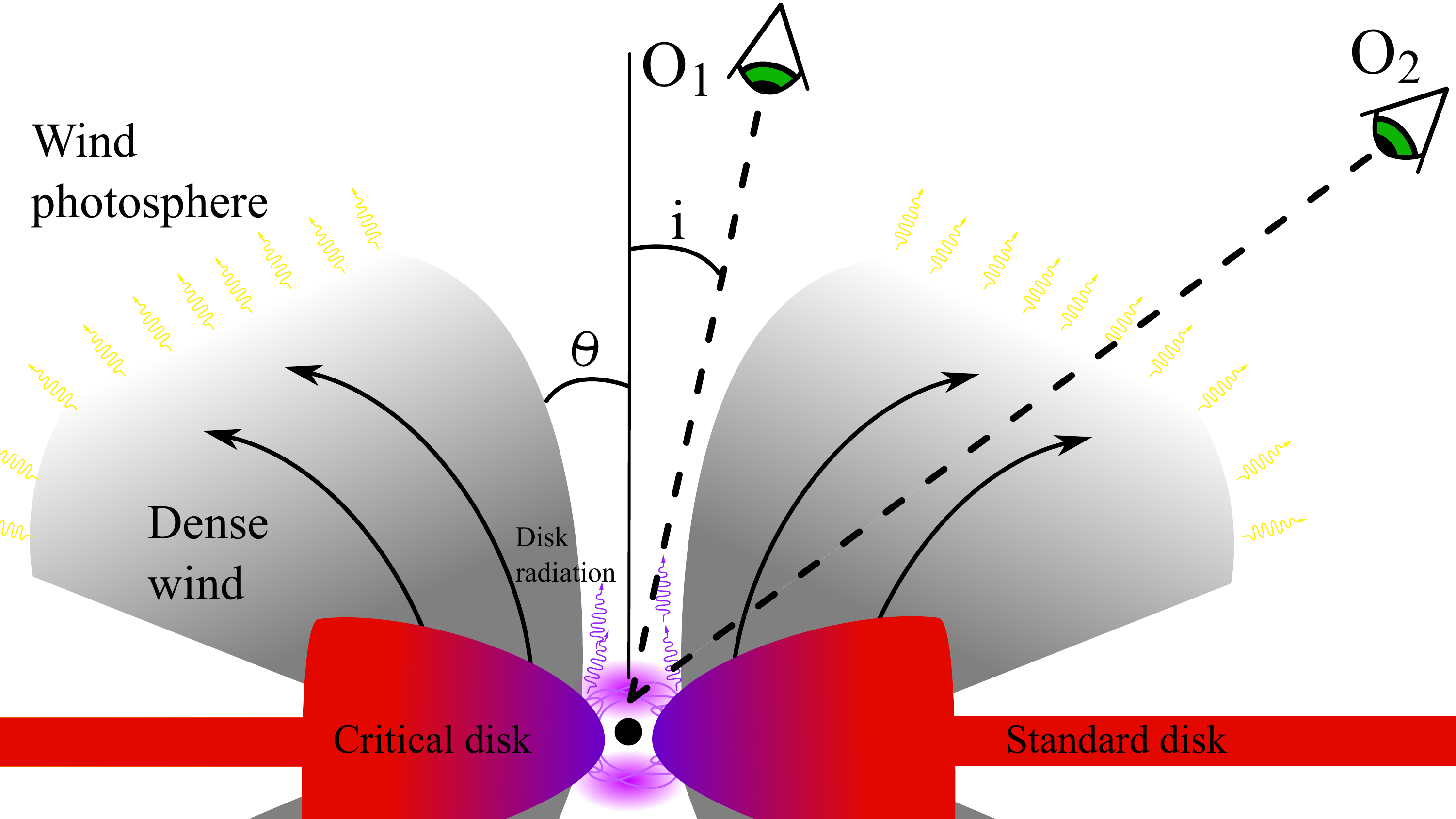}
    \caption{Schematic view of the environment around a super-Eddington AGN. The magnetically confined region is located in the immediate vicinity of the SMBH, where the inner disk provides an intense radiation field. The surrounding wind forms the dense walls of a funnel and can absorb high-energy radiation through both $\gamma\gamma$ interactions with the disk and wind photon fields and $\gamma N$ interactions with the outflowing matter. Depending on the inclination angle, the observer may directly view the inner disk through the funnel (observer $O_1$) or detect only reprocessed emission (observer $O_2$). Not to scale.}
    \label{fig:esquema_fuente}
\end{figure}


\begin{figure}
    \centering
    \includegraphics[width=0.9\linewidth]{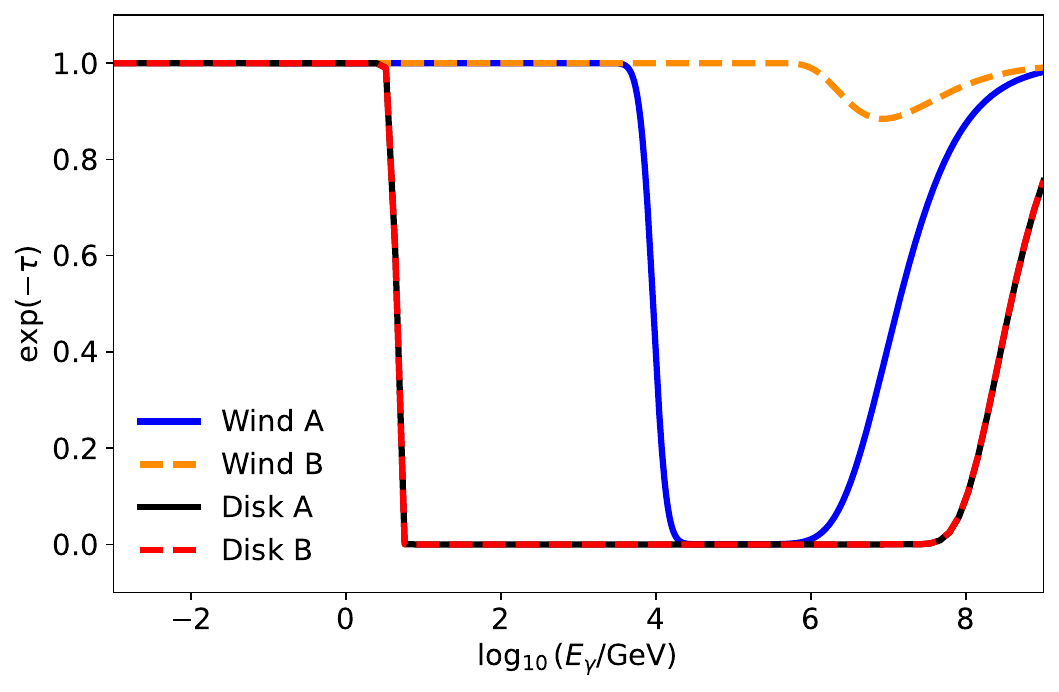}
\caption{Attenuation factor $\exp(-\tau_{\gamma\gamma})$ produced by $\gamma\gamma$ absorption in the thermal radiation fields of the wind photosphere and the inner disk, for models $A$ and $B$.}
    \label{fig:opacidad}
\end{figure}

The observed spectrum depends strongly on the geometry of the system and on the viewing angle $i$. A schematic representation is shown in Fig. \ref{fig:esquema_fuente}. In our scenario, the ejected wind forms the dense walls of a polar funnel that remains comparatively transparent to the thermal radiation from the inner disk. For lines of sight such that $i<\theta$ (observer $O_1$), where $\theta$ is the funnel semi-opening angle, the observer can directly view the innermost disk regions through the low-density channel. By contrast, for $i>\theta$ (observer $O_2$) the wind obscures the inner disk emission and efficiently absorbs a large fraction of the non-thermal radiation produced close to the SMBH. 

We first consider the absorption of gamma rays in the thermal radiation field of the wind photosphere. As discussed in Sec. \ref{sec:model}, the photosphere emits as a thermal source with temperature given by eq. \eqref{eq:T_field_wind}. Following the formalism of \citet{Abs_Cerutti_2011}, the corresponding $\gamma\gamma$ optical depth can be written as

\begin{equation}
\begin{aligned}
\tau_{\gamma\gamma,\rm wind}(E_\gamma)=
\int_{z_{\rm ph}}^{\infty} dl
\int_{\epsilon_{\rm thr}}^{\epsilon_{\max}} d\epsilon\,
&\sigma_{\gamma\gamma}(E_\gamma,\epsilon,\vartheta)\,
n_{\rm ph,wind}(\epsilon,R) \\
&\times \left(1-\cos\vartheta\right),
\end{aligned}
\label{eq:tau_gg_wind}
\end{equation}
where $z_{\rm ph}$ is the photospheric height, $\sigma_{\gamma\gamma}$ is the pair-production cross section, $n_{\rm ph,wind}$ is the differential photon density of the wind, and $\vartheta$ is the angle between the gamma-ray momentum and the target photon direction. The threshold energy is
\begin{equation}
\epsilon_{\rm thr}=\frac{2m_{\rm e}^{2}c^{4}}{E_\gamma(1-\cos\vartheta)}.
\end{equation}

If the system is observed at an inclination $i>\theta$, gamma rays can also be absorbed through interactions with matter in the wind. Since the mass outflow rate is of order $\dot M_{\rm w}\sim \dot m\,\dot M_{\rm Edd}$ and the wind extends over several hundred gravitational radii, the density around the central source can be high enough for photon-matter interactions to become relevant. Following \citet{DarkJets_Reynoso_Romero2008}, we compute the corresponding optical depth as
\begin{equation}
\tau_{\gamma N,\rm wind}(E_\gamma)=
\int_{0}^{\infty} dl\,
\sigma_{\gamma N}(E_\gamma)
\left(\frac{\rho_{\rm w}(R)}{m_{\rm p}}\right),
\label{eq:tau_gN_wind}
\end{equation}
where $\rho_{\rm w}(R)$ is the wind density given by eq. \eqref{ec: density_wind}, and $\sigma_{\gamma N}$ is the total cross section for high-energy photon-matter interactions, including pair creation and photomeson production channels. In our setup, this contribution becomes relevant already at MeV energies and leads to a strong suppression of the emission above $\sim1\,{\rm GeV}$.

We also account for $\gamma\gamma$ absorption in the radiation field of the innermost disk. For this purpose, we again follow the formalism of \citet{Abs_Cerutti_2011}, but considering a truncated disk geometry due to the onset of wind mass loss in the supercritical region. The total attenuation is then obtained by combining the effects of disk photons, wind photons, and wind matter with respect to the system's angle of inclination.


\begin{figure*}[h!]                            
\centering
  \centering
    \begin{subfigure}[t]{0.49\textwidth}
        \centering            \includegraphics[width=0.5\linewidth,trim= 130 30 120 20]{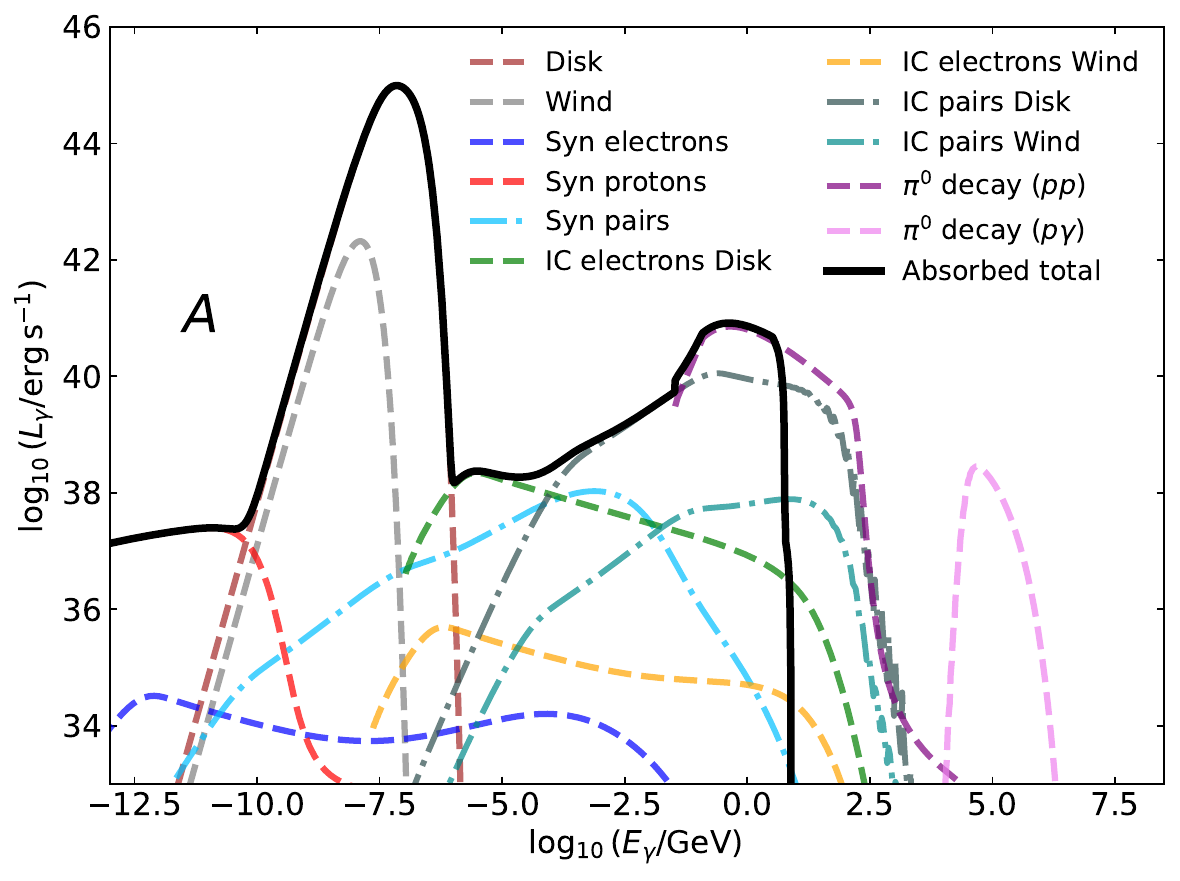} 
    \end{subfigure}
    \hfill
    \begin{subfigure}[t]{0.49\textwidth}
        \centering
    \includegraphics[width=0.49\linewidth,trim= 130 30 120 30]{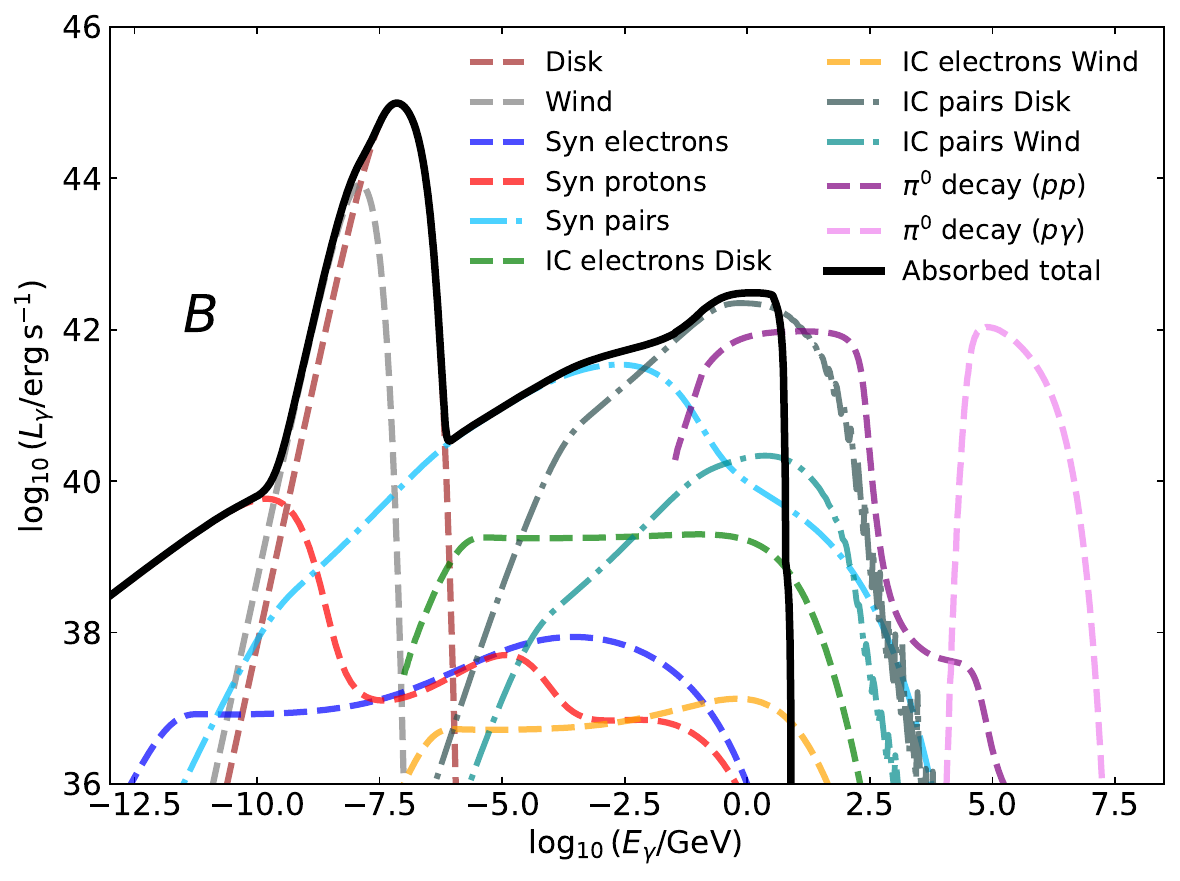} 
    \end{subfigure}
   
\caption{Spectral energy distributions for models $A$ and $B$, including the intrinsic radiative components and the total absorbed emission. The thermal contributions of the disk and the wind are also shown.}
 \label{fig:SEDS}
\end{figure*}


The resulting attenuation factors due to $\gamma\gamma$ absorption for a source viewed at low inclination, $i\approx5^\circ$, are shown in Fig. \ref{fig:opacidad}. Figure \ref{fig:SEDS} presents the SED, including both the intrinsic luminosity $L_\gamma$ and the absorbed luminosity. We also include the thermal contributions from the inner disk and from the wind.

At low energies, proton synchrotron radiation provides the dominant non-thermal contribution, since primary electrons cool efficiently through IC scattering in the intense disk radiation field (see Fig. \ref{fig:primary_rates_electron}). In the optical-UV range, the emission is dominated by the thermal components of the disk and wind. At higher energies, in the keV-MeV band, the radiation from secondary pairs becomes dominant, mainly through IC emission in model $A$ and through synchrotron radiation in model $B$. At GeV energies and above, the intrinsic gamma-ray emission is mainly produced by neutral-pion decay from both $pp$ and $p\gamma$ interactions, but this component is strongly attenuated by the large opacity of the surrounding radiation fields.

\section{Neutrino production} \label{sec:neutrino_flux}

Given the distributions of charged pions, $N_{\pi^\pm}(E_{\pi^\pm})$, muons, $N_{\mu^\pm}(E_{\mu^\pm})$, and kaons, $N_{K^\pm}(E_{K^\pm})$, we compute the emissivities of neutrinos and antineutrinos produced through their decay channels. Since, for the purposes of this work, neutrinos and antineutrinos are not distinguished observationally, we sum both contributions and hereafter refer to them simply as neutrinos.
We calculate the emissivity of main muon neutrinos produced by the decay of charged pions, $Q^{\pi^\pm}_{\nu_\mu}(E_\nu)$, as well as the emissivities of electron and muon neutrinos produced by muon decay, $Q^{\mu^\pm}_{\nu_e}(E_\nu)$ and $Q^{\mu^\pm}_{\nu_\mu}(E_\nu)$, including the helicity dependence of the muon channel. We also compute the emissivity of muon neutrinos from charged kaon decay, $Q^{K^\pm}_{\nu_\mu}(E_\nu)$, following the descriptions of \cite{Kelner_2006}, \citet{lipari2007}, \citet{Hummer_2010}, \citet{Reynoso_Romero_Magneticfield_2009}, and \citet{Neutrinos_Rey_Deus_2023}. The total neutrino emissivities are obtained by summing all the contributions, $Q_{\nu_e+\bar{\nu}_e}(E_\nu)$, and $Q_{\nu_\mu+\bar{\nu}_\mu}(E_\nu)$.



Assuming homogeneous and isotropic neutrino production within the interaction volume $\Delta V$, the unoscillated neutrino flux from a source at distance $d$ is written as
\begin{equation}
\phi^{(0)}_{\nu_{e,\mu}+\bar{\nu}_{e,\mu}}(E_\nu)=
\frac{\Delta V\,Q_{\nu_{e,\mu}+\bar{\nu}_{e,\mu}}(E_\nu)}{d^{2}}.
\label{eq:nu_unosc_flux}
\end{equation}
Since neutrino detection is most efficient in the muon-neutrino channel, owing to the long range of the secondary muons produced in neutrino telescopes, we focus on the muon-neutrino flux arriving at Earth. After flavor oscillations over astrophysical distances, this flux is given by
\begin{equation}
\phi_{\nu_\mu+\bar{\nu}_\mu}(E_\nu)=
\phi^{(0)}_{\nu_\mu+\bar{\nu}_\mu}(E_\nu)\,P_{\mu\mu}
+
\phi^{(0)}_{\nu_e+\bar{\nu}_e}(E_\nu)\,P_{e\mu},
\label{eq:nu_mu_earth}
\end{equation}
where $P_{\mu\mu}$ and $P_{e\mu}$ are the flavor-transition probabilities. These are computed as
\begin{equation}
P_{\mu\mu}=\sum_{i=1}^{3}|U_{\mu i}|^{2}|U_{\mu i}|^{2},
\qquad
P_{e\mu}=\sum_{i=1}^{3}|U_{e i}|^{2}|U_{\mu i}|^{2},
\label{eq:osc_prob}
\end{equation}
where $U$ is the neutrino mixing matrix, evaluated using the best-fit oscillation parameters reported by \citet{esteban2020}.

\begin{figure}
    \centering
    \includegraphics[width=0.9\linewidth]{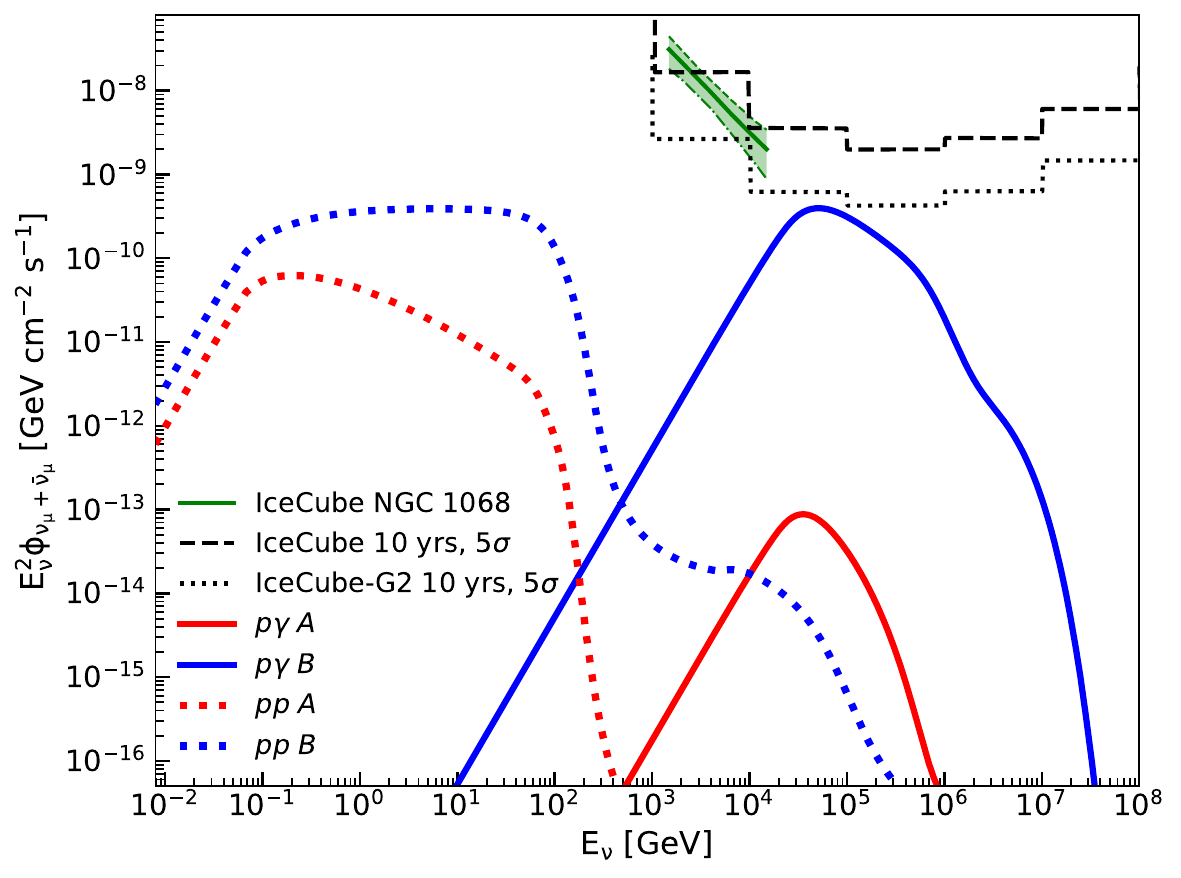}
    \caption{Muon-neutrino plus antineutrino flux for a super-Eddington AGN at $d=60\,{\rm Mpc}$, including the contributions from $pp$ and $p\gamma$ interactions for models $A$ and $B$. The 10-year, $5\sigma$ discovery potentials of IceCube and IceCube-Gen2 are also shown, together with the inferred flux of NGC 1068 for comparison.}
    \label{fig:neutrino_fluxes}
\end{figure}

The predicted neutrino flux from the magnetically confined region of a super-Eddington AGN located at $d=60$ Mpc is shown in Fig. \ref{fig:neutrino_fluxes}. In the figure we also include the sensitivities of IceCube and IceCube-Gen2 for 10 years of observation at the $5\sigma$ level and the inferred flux from NGC 1068 reported in \citet{IC_NGC1068_2022} for comparison. Red curves correspond to the total flux of model $A$, whereas blue curves correspond to model $B$.

The total neutrino flux is the composition of the neutrinos emitted by $pp$ as a plateau in the energy range of $10^{-1}$–$10^2$ GeV, and the neutrinos from the $p\gamma$ branch at higher energies between $10^5$–$3\times10^7$ GeV. The neutrino spectrum presents a peak at high energies in the $\sim 10$-$500$ TeV range. In model $A$, the flux reaches values of the order of $\sim 10^{-13}\,{\rm GeV\,cm^{-2}\,s^{-1}}$, while in model $B$ it reaches up to $\sim 3\times10^{-10}\,{\rm GeV\,cm^{-2}\,s^{-1}}$, where an second peak emerge at $\sim10$ PeV due $K^{\pm}$ decay. As expected from the particle distributions discussed in Sec. \ref{sec:cool_distro}, the $p\gamma$ channel dominates the neutrino output at the highest energies, whereas the $pp$ contribution is comparatively weaker. This behavior reflects the low matter density in the magnetically confined region and the strong radiative environment provided by the inner disk and the surrounding wind.

The expected number of muon-neutrino and antineutrino events per year in the energy range from $E_{\nu,\,\rm min}=2500\,{\rm GeV}$ to $E_{\nu,\,\rm max}=10^7\,{\rm GeV}$ is given by
\begin{equation}
\dot{\mathcal{N}}_{\nu_\mu+\bar{\nu}_\mu}= \int_{E_{\nu,\rm min}}^{E_{\nu,\,\rm max}}\mathrm{d}E_\nu \,A_{\nu,{\rm eff}}(E_\nu)\,\phi_\nu(E_\nu),
\end{equation}
where $A_{\nu,{\rm eff}}(E_\nu)$ is the effective area of the neutrino detector, which depends on the instrumental response, event selection, and analysis strategy. For IceCube, the observation of sources in the northern sky is favored because of the lower atmospheric background in the muon-track channel. This corresponds approximately to declinations $\delta>-15^\circ$, where the muon-track event selection can be efficiently applied. In Table \ref{table:events.60Mpc} we present the expected number of events per year for the IceCube \citep{IceCube_Gen2_2021,icecube2025}, the future upgrade IceCube-Gen2 \citep{ladneha_icecubegen2} and KM3NeT/ARCA \citep{adrianmartinez2016} observatories. These values represent expected event rates and should not be interpreted as directly reportable excesses. Establishing an excess requires a likelihood analysis against the atmospheric and diffuse astrophysical neutrino backgrounds; for reference, the IceCube evidence for NGC 1068 corresponds to $79^{+22}_{-20}$ excess events with a post-trial significance of $4.2\sigma$.

\begin{table}[h!]
{\small
\caption{Expected $\nu_\mu+\bar\nu_\mu$ neutrino event rates per year for a source at $d=60\,{\rm Mpc}$.}
\label{table:events.60Mpc}
\centering
\begin{tabular}{l c c c}
\hline
Model & IceCube & IceCube-Gen2 & KM3NeT \\
\hline
$A$ & $4.1\times 10^{-5}$ & $3.3\times 10^{-4}$ & $7.0\times 10^{-5}$ \\
$B$ & $2.3\times 10^{-1}$ & $1.8$ & $3.9\times 10^{-1}$ \\
\hline
\end{tabular}
}
\end{table}

These estimates show that model $B$ yields event rates that might become detectable with current or next-generation neutrino observatories, whereas model $A$ remains well below the sensitivity required for a plausible detection.

Therefore, while the gamma-ray output from the inner region is severely suppressed, neutrinos can escape essentially unaffected. In this sense, super-Eddington AGNs can naturally behave as hidden hadronic accelerators, with a strongly absorbed electromagnetic counterpart and a comparatively unobscured neutrino signal.

These results indicate that, under favorable conditions, super-Eddington Seyfert nuclei may produce neutrino fluxes within the reach of current and next-generation neutrino observatories. In particular, model $B$ lies much closer to the IceCube sensitivity and would be more readily testable with IceCube-Gen2, while model $A$ remains more challenging to detect. Therefore, the detectability of these systems is strongly controlled by the physical conditions in the innermost region, especially the level of magnetization and the resulting maximum proton energies.

\section{Application to NGC 7469} \label{sec:appl}

We apply our model to the super-accreting AGN NGC 7469. This nearby source has been extensively studied across the electromagnetic spectrum, from radio to X-rays, while no gamma-ray detections have been reported so far. \citet{SMBH_Du_2015} classified NGC 7469 as a super-Eddington accreting massive black hole, although the reported accretion rate, $\log(\dot M)=0.9^{+1.83}_{-1.87}$, remains highly uncertain. The SMBH mass is estimated to lie in the range $M_{\rm BH}=9\times10^{6}$-$10^{7}\,M_\odot$, and the source is located at a distance of $D\approx70$ Mpc. Although the inclination of the system is not tightly constrained, low disk inclinations of $\lesssim20^\circ$ have been inferred \citep{opticngc7469_Prince_2025}. 

NGC 7469 has been observed with VLBI, where its weak nuclear radio emission reveals a faint core jet-like structure on scales of $\sim100$ pc, together with extended diffuse radio emission associated with the surrounding star-forming region \citep{VLBIngc7469_Lonsdale_2003,Radiongc7469_Orienti_2010}. JWST infrared observations have revealed active star formation in a circumnuclear ring, as well as a decelerating, stratified outflow emerging from the nucleus, with a mass-loss rate of $\sim1$-$5\,M_\odot\,{\rm yr^{-1}}$ and an inferred nuclear inflow rate of $\dot M=0.2$-$17\,M_\odot\,{\rm yr^{-1}}$ \citep{IRngc7469_Lai_2022,IRngc7469_Armus_2023,IRngc7469_Bianchin_2024}. The source has also been extensively studied in the optical and UV bands through reverberation mapping, yielding a central black-hole mass of order $\sim10^{7}\,M_\odot$ \citep{opticngc7469_peterson_2014,opticngc7469_Prince_2025,opticalngc7469_saule_2025}. In X-rays, NGC 7469 exhibits strong variability and a complex spectrum, with a high-energy cutoff around 170 keV, a soft excess, and a continuum commonly modeled through Comptonization in a hot-warm corona \citep{xrayngc7469_markowitz_2010,xrayngc7469_middei_2018,xrayngc7469_Pahari_2020,xrayngc7469_Partington_2025,xrayngc7469_Feuillet_2026}. No significant gamma-ray detection has been reported, and \citet{neutrinongc7469_Yang_2025} found no evidence for GeV emission in their analysis of \textit{Fermi} data.

NGC 7469 has recently emerged as a promising candidate high-energy neutrino emitter. Excluding NGC 1068, this source was reported as the most significant object in the IceCube search over 47 X-ray-bright non-blazar AGNs \citep{IC_Xrays_2026}, with a global significance of $2.4\sigma$ in the survey of hard X-ray emitters. The reported neutrino excess is mainly associated with the two events IC 220424A and IC 230416A, with energies of $\sim184$ TeV and $\sim127$ TeV, respectively \citep{neutrinongc7469_Yang_2025}, and the best-fit neutrino spectrum has a hard spectral index of $\approx1.9$. In addition, the best-fit neutrino flux and its $1\sigma$ uncertainty pose difficulties for standard core-corona interpretations \citep{neutirnosngc7469_sommani_2025,IC_Xrays_2026}. \citet{neutrinongc7469_Yang_2025} modeled the neutrino emission within a coronal scenario in which the source properties are constrained by observations and the neutrino flux arises from $p\gamma$ interactions with the X-ray coronal radiation field. By contrast, our model starts from a basic super-Eddington disk configuration and assumes that neutrinos are produced through interactions with the thermal photon field of the disk and its associated wind.

In modeling the source, we adopt the parameter set listed in Table \ref{tab:model_parameters_ngc7469}. Variations in basic disk parameters such as the viscosity $\alpha$ or the adiabatic index $\gamma$ produce only minor changes in the disk temperature and magnetic field strength, whereas the most relevant quantity is the disk magnetization parameter $\beta_{\rm disk}$. Since the neutrino data suggest a hard spectral index of $1.9$, we also explore a lower-density acceleration region with stronger magnetization ($\sigma_{\rm gas}$), corresponding to an injected proton index of $\Gamma=1.5$, which results in a harder and more intense neutrino spectrum.

\begin{figure*}[t]
    \centering
    \includegraphics[width=0.8\textwidth]{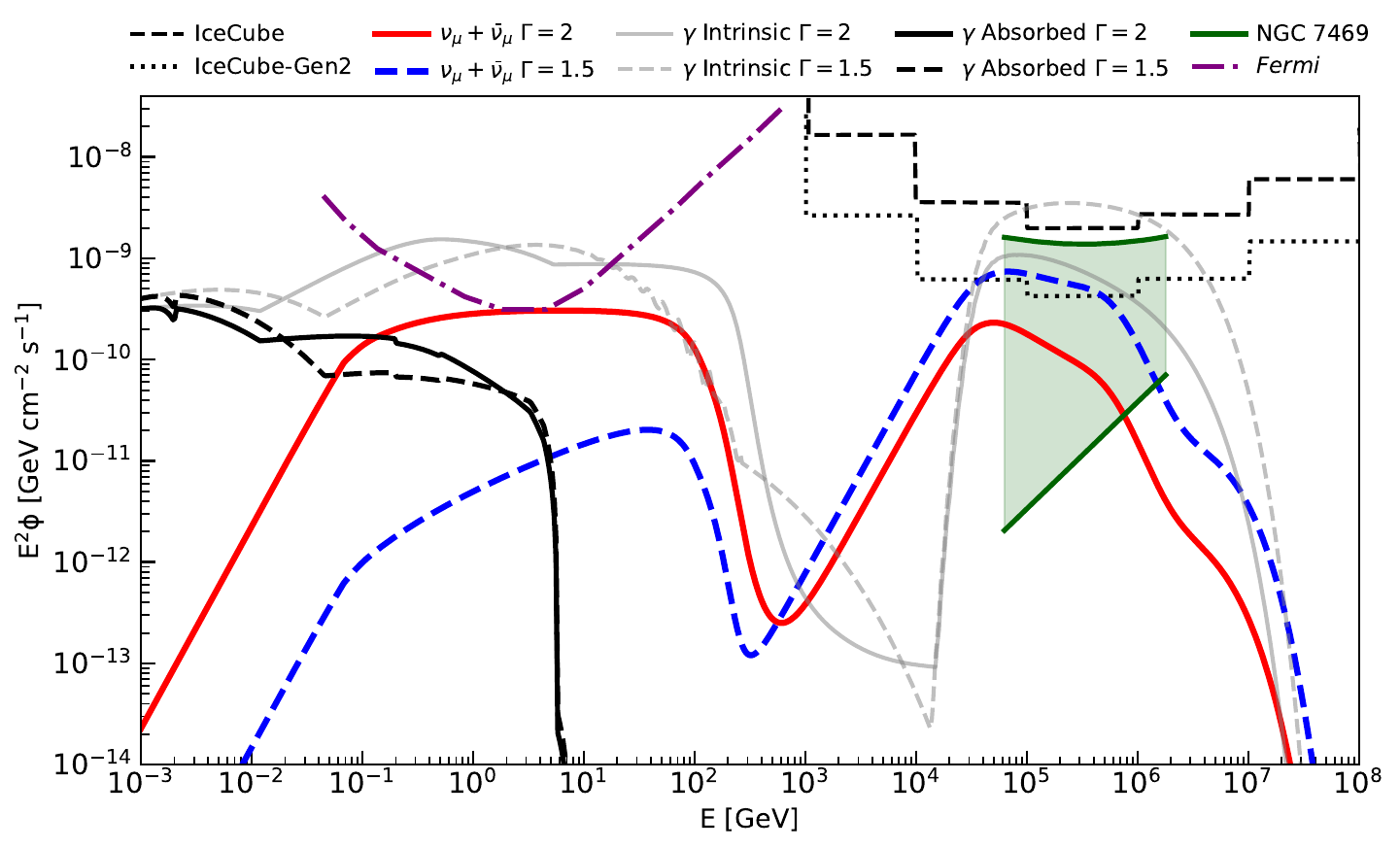}
    \caption{Predicted non-thermal electromagnetic radiation and muon-neutrino flux from NGC 7469, for proton injection indices of $\Gamma=2$ (solid lines) and $\Gamma=1.5$ (dashed lines). The grey lines show the total intrinsic electromagnetic emission, while the black lines represent the absorbed radiation. The blue and red lines correspond to the muon-neutrino flux, which is unaffected by absorption. We also show the 10-year, $5\sigma$ sensitivity curves of IceCube and IceCube-Gen2, together with the gamma-ray sensitivity of \textit{Fermi} and the best-fit neutrino flux of NGC 7469 inferred by IceCube.}
\label{fig:ngc7469_flux}
\end{figure*}

In Fig.~\ref{fig:ngc7469_flux} we present the predicted non-thermal electromagnetic and neutrino fluxes for NGC 7469 that results from the application of our model with the mentioned parameters. We also show the 10-year, $5\sigma$ sensitivity curves of IceCube and IceCube-Gen2, together with the gamma-ray sensitivity of \textit{Fermi}. In addition, we include the best-fit all-flavor neutrino flux constrained in the 95\% C.L. constrained energy range and its $1\sigma$ uncertainty reported for NGC 7469 by \citet{IC_Xrays_2026}. We take into account all relevant absorption processes affecting the high-energy radiation, finding that $\gamma\gamma$ absorption with disk photons completely suppresses the gamma-ray emission above $\sim10$ GeV. The neutrino spectrum peaks at approximately the same energy as in models $A$ and $B$, namely at $E_\nu\approx0.4$ PeV, reaching a flux of $\approx2$-$8\times10^{-10}\,{\rm GeV\,cm^{-2}\,s^{-1}}$ depending on the hardness of the injected proton distribution. Within this framework, the model can account for the reported neutrino events from NGC 7469 based on its properties as a super-accreting source.

\begin{table}[t]
\centering
\small
\setlength{\tabcolsep}{3pt} 
\caption{Values adopted for NGC 7469. Values derived from observations and other estimates are marked with $^\star$, assumptions with $^\oplus$, and derived quantities with $^\otimes$.}
\label{tab:model_parameters_ngc7469}

\begin{tabularx}{\columnwidth}{>{\raggedright\arraybackslash}X c c c}
\hline
\textbf{Parameter (symbol)} & Case $\Gamma=2$ & Case $\Gamma=1.5$ & Units \\
\hline
BH mass$^\star$ ($M_{\rm SMBH}$) 
& $10^{7}$ & $10^{7}$ & $M_\odot$ \\

Distance$^\star$ ($d$) 
& $70$ & $70$ & Mpc \\

System inclination$^\star$ ($i$) 
& $15$ & $15$ & $^\circ$ \\

Accretion rate$^\star$ ($\dot m$) 
& $10$ & $10$ & $\dot M/\dot M_{\rm Edd}$ \\

Disk magnetization$^\oplus$ ($\beta_{\rm disk}$) 
& $2$ & $2$ &  \\

Disk opening angle$^\otimes$ ($\delta$) 
& $20^\circ$ & $20^\circ$ &  \\

Magnetic field$^\otimes$ ($B$) 
& $2.6\times10^{4}$ & $2.6\times10^{4}$ & $\mathrm{G}$ \\

Spectral index$^\oplus$ ($\Gamma$) 
& $2$ & $1.5$ &  \\

Magnetic power$^\otimes$ ($L_{\rm mag}$) 
& $4\times10^{44}$ & $4\times10^{44}$ & $\mathrm{erg\,s^{-1}}$ \\

Number density$^\otimes$ ($n_i$) 
& $4.8\times10^{9}$ & $9.5\times10^{8}$ & $\mathrm{cm^{-3}}$ \\

Volume acc. region$^\otimes$ ($\Delta V$) 
& $1.5\times10^{38}$ & $1.5\times10^{38}$ & $\mathrm{cm^{3}}$ \\

Max. proton energy$^\otimes$ ($E^{\max}_{\rm p}$) 
& $2\times10^{16}$ & $2\times10^{16}$ & $\mathrm{eV}$ \\
\hline
\end{tabularx}
\end{table}

\section{Discussion}\label{sec:discussion}

Models $A$ and $B$ both correspond to super-Eddington accretion regimes, differing mainly in the degree of magnetization of the inner disk and acceleration region. For a fixed black-hole mass, the disk luminosity remains regulated around the Eddington value and therefore does not vary dramatically with the accretion rate, whereas the magnetic field strength is more sensitive to the magnetization parameter $\beta_{\rm disk}$ and to the detailed structure of the inner flow. Since most Seyfert galaxies are radio-quiet systems, whereas only a small fraction are radio-loud or exhibit clear jet activity, low or moderate magnetization levels are expected to be more representative of the population as a whole. In this context, a higher magnetization mainly leads to a harder non-thermal particle distribution and, consequently, to a neutrino spectrum extending to higher energies.

An important requirement of the model is that the magnetically confined region must remain sufficiently transparent to be consistent with the observation of thermal emission from the inner disk. In our scenario, this region is continuously replenished while matter is also removed through diffusion or convective transport, allowing a quasi-steady state to be maintained. Owing to its comparatively low density, the population of relativistic particles escaping from the acceleration zone is not expected to undergo strong hadronic interactions in the surrounding outflow. This is reflected in the interaction rates and in the relatively weak neutrino contribution from the $pp$ channel at high energies. At the same time, the large population of lower-energy protons can still produce a non-negligible gamma-ray component through the decay of neutral pions generated in $pp$ collisions.

Although neutral and charged pions are produced in comparable amounts in hadronic interactions, the resulting gamma-ray and neutrino outputs do not necessarily remain comparable. Charged pions, muons, and kaons are subject to radiative losses before decaying, which modify the neutrino spectrum at the highest energies. The gamma-ray component associated with neutral-pion decay, in turn, is strongly attenuated by $\gamma\gamma$ absorption in the intense radiation fields of the disk and wind, and by $\gamma N$ interactions at high inclination angles. The source can therefore appear bright in neutrinos while remaining faint or undetected at very high gamma-ray energies.

A distinctive feature of the present model is its implication for the low-energy non-thermal emission. As shown by the SEDs in Fig.~\ref{fig:SEDS}, synchrotron radiation from protons dominates over that from primary electrons. This results from the strong IC cooling experienced by electrons, which substantially reduces their synchrotron output. As a result, although proton synchrotron emission is intrinsically much less efficient, it becomes the dominant non-thermal contribution in the radio band. This may provide a potentially useful observational signature of highly radiative, proton-dominated inner environments.

In this context, NGC 1068 provides an interesting observational benchmark. This source, located at $d\approx 14.4$ Mpc, hosts a SMBH with an estimated mass in the range $M_{\rm BH}\sim 10^{7}\,M_\odot$ and is usually classified as an obscured Seyfert 2 galaxy. Its neutrino signal is commonly interpreted within sub-Eddington scenarios, in which the innermost accretion flow is associated with a hot X-ray-emitting corona. Our model differs substantially from that picture, since it assumes a supercritical accretion flow with a geometrically thick inner disk and a dense radiation-driven wind. Within our scenario, the neutrino spectrum is shifted toward higher energies, with a broad maximum around $\sim 100$ TeV, potentially helping to distinguish super-Eddington hidden cores from more standard coronal scenarios if future neutrino statistics improve.

TDEs provide a natural transient extension of the super-Eddington accretion scenario discussed here. Even galaxies hosting sub-Eddington or moderately accreting SMBHs can undergo short episodes in which a star is disrupted by the tidal field of the SMBH. A fraction of the bound debris returns to the central region, circularizes, and can temporarily drive the accretion flow to highly super-Eddington rates, with peak fallback rates reaching tens to hundreds of $\dot M_{\rm Edd}$ for $M_{\rm BH}\sim10^{6}\,M_\odot$ \citep{Review_TDE_Gezari_2021,TDE_multi_Wevers_2023}. Early estimates of the per-galaxy TDE rate are typically $\sim10^{-5}$--$10^{-4}\,{\rm yr^{-1}}$, although it may be enhanced up to $\sim10^{-3}\,{\rm yr^{-1}}$ in dense nuclear environments \citep{TDE_Kaur_2025}. Optical--UV studies report a volumetric TDE rate of $\sim7\times10^{-7}\,{\rm Mpc^{-3}\,yr^{-1}}$ for SMBHs in the mass range $M_{\rm SMBH}=10^{5.5}$--$10^{7.5}\,M_\odot$ \citep{TDE_Velzen_2018,TDE_Velzen_2020}, while \citet{TDE_neutrinos_Wu_2024} estimate a rate of $\sim3\times10^{-6}\,{\rm Mpc^{-3}\,yr^{-1}}$ using theoretical dynamical models. For a sphere of radius $100\,{\rm Mpc}$, corresponding to a volume $V=4.19\times10^{6}\,{\rm Mpc^{3}}$, these values imply a TDE rate of $\sim2.9$--$12.5\,{\rm yr^{-1}}$. Since only a small fraction of TDEs are expected to launch relativistic jets, this population is broadly consistent with the super-Eddington accretion scenario without jets explored in this work \citep{TDE_Stone_2016}.

This connection is particularly relevant for AGN, where the presence of a pre-existing accretion disk may enhance the TDE rate and modify the observable flare properties \citep{TDE_Seyfert_Somalwar_2023,Nine_TDE_Ferris_2025,TDE_sample_Jiang_2025}. Several TDE candidates have been reported in NLS1 or AGN-like environments, including PS16dtm, AT2019fdr, and the repeating partial TDE candidate AT2021aeuk \citep{TDE_Blanchard_2017,Neutrino_TDE_Reusch_2022,TDE_NLS1_Sun_2025}. In such systems, however, distinguishing genuine TDEs from extreme accretion-disk variability remains observationally challenging.

TDEs have also emerged as promising candidate sources of high-energy neutrinos and ultrahigh-energy cosmic rays \citep{neutrino_TDE_Jiang_2023,TDE_multi_Wevers_2023,TDE_neutrinos_Wu_2024,TDE_neutrinos_Velzen_2024,CR_neutrinos_TDE_Plotko_2025,TDE_neutrino_Langis_2026}. High-energy neutrinos have been tentatively associated with TDEs and TDE-like accretion flares, including the well-studied association AT2019dsg--IC191001A, with a neutrino energy of $\approx0.2\,{\rm PeV}$ and a source distance of $D\approx230\,{\rm Mpc}$, as well as other possible coincidences such as AT2019fdr--IC200530A, AT2019aalc--IC191119A, and the obscured radio-emitting TDE candidate SDSS J151345.75+311125.2--IC170514B. The reported neutrino energies, ranging from $\sim 80\,{\rm TeV}$ to $\sim 0.2\,{\rm PeV}$ \citep{TDE_neutrino_Stein_2021,Neutrino_TDE_Reusch_2022,TDE_neutrinos_Velzen_2024,TDE_neutrino_Zhou_2026}, are comparable to the characteristic neutrino energies predicted in our model (see Fig.~\ref{fig:neutrino_fluxes}). Although the statistical significance of these associations and the underlying physical mechanism remain under debate, they suggest that transient super-Eddington accretion episodes in AGN-like environments may contribute to the observed high-energy neutrino sky.

The model of \citet{neutrinongc7469_Yang_2025} relies on a standard coronal configuration, in which protons are accelerated up to PeV energies and interact predominantly with the coronal X-ray photon field. By contrast, our framework explicitly assumes a super-Eddington accretion regime, $\dot m\gtrsim 1$, in which the accretion flow becomes naturally optically and geometrically thick and launches powerful radiation-driven winds. This difference modifies both the target photon fields and the relevant opacity scales. Instead of a compact coronal plasma, we locate particle acceleration in a magnetically confined disk funnel, where relativistic hadrons interact mainly with thermal from the supercritical disk. The existence of this funnel is supported by the X-ray detections of the source. Another important distinction concerns the suppression of the accompanying gamma-ray emission. In the coronal model of \citet{neutrinongc7469_Yang_2025}, the size of the emitting region is constrained by the $Fermi$ upper limits. In our model, the disk and dense wind walls naturally provide a large optical depth, so that the gamma rays produced together with neutrinos are absorbed and reprocessed through $\gamma\gamma$ and $\gamma N$ interactions. This offers a physically motivated mechanism for hiding the high-energy electromagnetic counterpart while preserving an escaping neutrino signal. Therefore, a hard neutrino spectrum such as that inferred for NGC 7469, with a spectral index of $\sim1.9$, could in principle be accommodated in a super-Eddington hidden-core scenario with a hard hadronic injection index, $\Gamma\simeq1.5$-$2$. 

\section{Conclusions} \label{sec:conclusion}

We have investigated neutrino production in super-Eddington AGN under conditions of low and moderate magnetization. Our results show that these systems can operate as efficient hadronic accelerators, with protons reaching energies of $\sim 1$--$10$ PeV depending on the magnetic field strength near the SMBH. Because of the intense photon field provided by the supercritical disk and the surrounding wind, the dominant hadronic interactions occur in a highly radiative environment.

Within this framework, electrons and secondary pairs cool efficiently through IC processes, giving rise to a significant keV--GeV radiative component, while relativistic protons produce secondary pairs, high-energy photons, and neutrinos. At the same time, radiative losses suffered by pions, muons, and kaons modify the final neutrino output at the highest energies. The resulting neutrino spectrum peaks in the $\sim 10$--$100$ TeV range, with the most optimistic scenarios yielding fluxes that may approach the sensitivity limits of current and next-generation neutrino observatories.

A key feature of the model is that the inner environment of a super-Eddington system is dense in both radiation and matter. As a consequence, although neutrinos can escape from the central region essentially unaffected, the accompanying high-energy electromagnetic emission is strongly attenuated through $\gamma\gamma$ and $\gamma N$ absorption, depending on the inclination angle of the system. Super-Eddington AGN may therefore behave as hidden hadronic accelerators, in which efficient neutrino production coexists with a faint or strongly reprocessed gamma-ray counterpart.

We also applied the model to the nearby super-Eddington Seyfert galaxy NGC 7469. For a moderately magnetized configuration, the predicted neutrino flux reaches a maximum around $\sim 0.4\,{\rm PeV}$ is compatible with the flux reported by IceCube, whereas the associated gamma-ray emission is strongly suppressed by the dense environment. This source therefore provides a representative example of a radio-quiet Seyfert nucleus in which supercritical accretion, strong winds, and hidden hadronic activity may coexist.

Our results support the idea that super-Eddington Seyfert nuclei may contribute to the population of neutrino-emitting AGN, particularly when the gamma-ray emission from the central engine is efficiently absorbed and reprocessed. Transient super-Eddington episodes, such as those triggered by tidal disruption events in AGN-like environments, provide a natural channel for realizing these conditions. Future searches combining high-energy neutrino observations with optical, UV, X-ray, and gamma-ray data will be crucial for identifying these systems and constraining the physics of supercritical accretion flows and hidden AGN cores.

\section*{Acknowledgements}
GER was funded by PID2022-136828NB-C41/AEI/10.13039/501100011033/ and through the ``Unit of Excellence María de Maeztu'' award to the Institute of Cosmos Sciences (CEX2019-000918-M, CEX2024-001451-M). Additional support came from PIP 0554 (CONICET). MMR acknowledges support from UNMdP through grant 80020240500217MP.

\appendix


\bibliographystyle{elsarticle-harv} 
\bibliography{example}






\end{document}